\documentclass[a4paper,11pt]{article}
\usepackage{jcappub} 
\usepackage{enumitem}
\usepackage{bm}
\usepackage{soul}
\usepackage{comment}
\newcommand{\Pygwb}{\texttt{Pygwb}}
\newcommand{\Bilby}{\texttt{Bilby}}

\title{\boldmath 
Forecast Analysis of Astrophysical Stochastic Gravitational Wave Background Beyond General Relativity: A Case Study on Brans-Dicke Gravity}

\author[a,b]{Ran Chen,}
\author[b,c,d]{Zhao Li,}
\author[a]{Yin-Jie Li,}
\author[a]{Yi-Ying Wang,}
\author[b,c]{Rui Niu,}
\author[b,c]{Wen Zhao,}
\author[a,b]{and Yi-Zhong Fan\footnote{Corresponding Author}}
\affiliation[a]{Key Laboratory of Dark Matter and Space Astronomy, Purple Mountain Observatory, Chinese Academy of Sciences, Nanjing 210033, People's Republic of China}
\affiliation[b]{School of Astronomy and Space Sciences, University of Science and Technology of China, Hefei 230026, People's Republic of China}
\affiliation[c]{CAS Key Laboratory for Researches in Galaxies and Cosmology, Department of Astronomy, University of Science and Technology of China, Chinese Academy of Sciences, Hefei, Anhui 230026, People's Republic of China}
\affiliation[d]{Department of Physics, Kyoto University, Kyoto 606-8502, Japan}

\emailAdd{ranchen@pmo.ac.cn, lz111301@mail.ustc.edu.cn, liyinjie@pmo.ac.cn, wangyy@pmo.ac.cn, nrui@ustc.edu.cn, wzhao7@ustc.edu.cn, yzfan@pmo.ac.cn}

\abstract{
Scalar-tensor gravity, exemplified by Brans-Dicke (BD) gravity, introduces additional scalar polarization modes that contribute scalar radiation alongside tensor modes.
We conduct a comprehensive analysis of how gravitational wave generation and propagation effects under Brans-Dicke gravity are encoded into the astrophysical stochastic gravitational wave background (AGWB).
We perform end-to-end analyses of realistic populations of simulated coalescing binary systems to generate AGWB mock data with third-generation gravitational wave detectors and conducted a complete Bayesian analysis for the first time.
We find the uncertainties in the population properties of binary black holes (BBH) significantly affect the ability to constrain BD gravity. 
Furthermore, we explore the detectability of potential scalar backgrounds that originates from binary neutron star (BNS) and neutron-star–black-hole (NSBH) mergers, with NSBH systems expected to modify the spectral index of the scalar background and introduce oscillatory behavior.
We show that the observations of the AGWB enable the separation of mixed tensor and scalar polarization modes with comparable sensitivity to each mode. 
However, the scalar background is expected to remain substantially weaker than the tensor background, even in scenarios where BD gravity exhibits significant deviations from general relativity (GR), resulting only upper limits can be placed on the scalar background.
We conclude that for ambiguous populations, employing waveform matching with individual sources provides a more robust approach to constrain BD gravity.}

\begin{document}
\maketitle
\flushbottom

\section{Introduction}
\label{sec:intro}
As one of the most significant predictions of general relativity (GR), gravitational waves (GWs) open new avenues for exploring the nature of gravity and cosmology~\cite{LIGOScientific:2016aoc,LIGOScientific:2016sjg,LIGOScientific:2017vwq,LIGOScientific:2017zic,LIGOScientific:2018mvr,LIGOScientific:2020ibl,KAGRA:2021vkt} after the successful detection from individual compact binary sources over the past few years.
These confirmed detections are verified by matching the resolved waveforms generated by individual and point-like sources to the detector data streams, which constitute only a tiny fraction of the gravitational-wave sky. 
In addition, the stochastic GW backgrounds (SGWBs) ~\cite{Regimbau:2011rp,Thrane:2013oya,Christensen:2018iqi,Renzini:2022alw} are generated by multiple point sources or extended sources, presenting as an incoherent superposition of a vast collection of unresolved signals.
In a recent series of observations, multiple Pulsar Timing Array collaborations, including NANOGrav~\cite{NANOGrav:2023gor, NANOGrav:2023hvm}, EPTA~\cite{EPTA:2023fyk}, and CPTA~\cite{Xu:2023wog}, have reported evidence for nanohertz SGWBs.
While alternative cosmological models have been proposed to interpret the nanohertz SGWBs.
The most conservative interpretation is astrophysical, attributing the  signal to supermassive black hole binaries~\cite{NANOGrav:2023hfp,EPTA:2023gyr,Ellis:2023dgf,Ellis:2023oxs}.

Astrophysical stochastic gravitational-wave backgrounds (AGWBs) in the ground-based interferometer band originate from distant compact binaries with low signal-to-noise ratios (SNRs) and are expected to be confirmed by ground-based gravitational-wave detector networks~\cite{Rosado:2011kv,Zhu:2011bd} in the near future.
AGWBs serve as a window for probing various questions concerning the nature of gravity, the population properties of compact binaries, and cosmology.
However, the stochastic backgrounds, the scalar, and the vector modes predicted by alternative theories of gravity have not been detected in the third LIGO-Virgo-KAGRA collaboration observing run (O3)~\cite{KAGRA:2021kbb}.
Based on their enhanced sensitivity and angular resolution, the next-generation ground-based detectors, such as the Cosmic Explorer (CE)~\cite{Reitze:2019iox}, Einstein Telescope (ET)~\cite{Maggiore:2019uih}, will have the potential to detect the AGWB and extract valuable information about gravity and cosmology~\cite{Chen:2024xkv,Nishizawa:2019rra,Takeda:2019gwk}.

Although GR is widely acknowledged as the most successful theory of gravitation, there is various theoretical and experimental evidence that challenges this standard model~\cite{Addazi:2021xuf,Dolgov:2003px,Clifton:2011jh,Philcox:2022hkh,Hou:2022wfj}.
Theories of gravity beyond GR have predicted several characteristic features in GWs,
including energy loss, polarizations, dispersion, speed, and others. 
The direct constraints on the contributions of these extra polarizations to the SGWB have been investigated in~\cite{LIGOScientific:2018czr}. 
The revisions of the GWs in GR by these features can vary the spectral shape of the AGWB. 
Several modified gravity theories~\cite{Maselli:2016ekw,Saffer:2020xsw} have been analyzed by the deviations of the expected signal of the AGWB. 
The velocity birefringence effects in AGWB have been used to place constraints on parity-violating theory~\cite{Callister:2023tws}.

Incorporating an additional scalar field, scalar-tensor theories are always employed to explain the late-time acceleration of the Universe~\cite{Brax:2004qh, Baccigalupi:2000je, Riazuelo:2001mg,Laya:2022vns} or cosmic inflation~\cite{Clifton:2011jh, Barrow:1990nv}. 
These theories have garnered considerable attention in the literature over recent years~\cite{Rossi:2019lgt,SolaPeracaula:2020vpg,Kramer:2021jcw,Errehymy:2022hxt}. 
As the simplest and most extensively studied ones, BD theory describes the curvature of spacetime, where the Ricci scalar is non-minimally coupled to a massless scalar field that serves as Newton's gravitational constant~\cite{Brans:1961sx}.
The extra degree of freedom introduced by the massless scalar field results in an additional ``breathing" polarization of GWs and accompanying scalar radiation.
Using the observations of the Shapiro time delay~\cite{Postnov:2014tza,Bertotti:2003rm}, the Cassini mission's measurement of the parameterized post-Newtonian parameter provided the most stringent result of the coupling constant $\omega_{\rm BD}$ with $\omega_{\rm BD} > 40000$ at the 2$\sigma$ level.
On larger scales, using Cosmic Microwave Background data from Planck, the BD parameter $\omega_{\rm BD}$ was constrained to $\omega_{\rm BD} > 692$ at the 99\% confidence level~\cite{Avilez:2013dxa}.
The prospect constrain of $\omega_{\rm BD}$ has been investigated by the Fisher matrix method with future space-based gravitational wave detectors, specifically the Laser Interferometer Space Antenna (LISA)~\cite{Yagi:2009zm}.
Based on the real gravitational wave signals from binary compact star coalescences, a joint analysis of GW200115 and GW190426\_152155 constrained $\omega_{\rm BD} > 40$ at the 90\% confidence level~\cite{Niu:2021nic}.
Analyzing the GW200115 data, an observational constraint of $\omega_{\rm BD} > 81$ by exploiting a waveform for a mixture of tensor and scalar polarizations~\cite{Takeda:2023wqn}.
Additionally, by incorporating the dominant (2, 2) mode correction and higher harmonic revisions, similar results were obtained with $\omega_{\rm BD} > 5$ at the 90\% level~\cite{Tan:2023fyl}.

Therefore, it remains meaningful to investigate the future constraints of $\omega_{\rm BD}$ based on the AGWB signals. Because the AGWB represents the integrating signals of various sources in the universe, it concerns wide spatial and temporal scales. 
Moreover, the signals of the AGWB indicate the response of non-tensor modes, providing the possibility to test specific gravitational models from an independent perspective. 
As a result, a particularly suitable candidate for such investigations is BD theory. 

Unlike previous studies that employ power-law models to search for the SGWB, our work comprehensively accounts for waveform corrections during the stages of the generation and propagation of GW for binary compact star systems. 
Under BD gravity, these corrections can be integrated into the spectrum of the AGWB.
Based on the third-generation gravitational wave detectors, we explore the prospective constraints of BD gravity by end-to-end analyses of AGWB simulations with fully specified parent BBH populations. Furthermore, we prospect the potential of detecting GW scalar polarization modes that are produced by BNS and NSBH systems.

This paper is organized as follows.
In Section~\ref{sec:Stochastic Gravitational-Wave Background}, we outline the methodology for calculating the SGWB spectrum in BD gravity, incorporating the effects of GW generation and propagation. 
Additionally, we introduce the tools employed for detecting the stochastic background, extending the analysis to BD gravity with extra scalar polarization.
In Section~\ref{sec:Simulation and Parameter Estimate}, we describe the methodology used for generating the simulated datasets and discuss the results of Bayesian parameter estimation from two perspectives: a full analysis based on waveforms and the search for scalar modes.
In Section~\ref{sec:Conclutions}, we provide our conclusions and discussion. 
Throughout this work, we adopt the metric signature $(-, +, +, +)$ and utilize the unit convention $c = 1$. 
In our work, we adopt the result of Planck18~\cite{Planck:2018vyg} for the value of cosmology parameters, Hubble constant $H_0 = 67.66~\mathrm{km}~\mathrm{s}^{-1} \mathrm{Mpc}^{-1}$, the matter density parameter $\Omega_{m}=30.97\times10^{-2}$ and the cosmological constant density parameter $\Omega_{\Lambda}=68.88\times10^{-2}$.

\section{Theoretical Methodology}
\label{sec:Stochastic Gravitational-Wave Background}

\subsection{SGWB spectrum in GR}
\label{sec:Energy density of background}
The quantity of interest in stochastic searches is usually chosen to be the SGWB spectrum, defined as \cite{Allen:1997ad},
\begin{equation}\label{omg_def}
\Omega_{\mathrm{GW}}(f) \equiv \frac{1}{\rho_{c}} \frac{\mathrm{d} \rho_{\mathrm{GW}}}{\mathrm{d} \ln f},
\end{equation}
where $\rho_{c}\equiv3H_0^2/8\pi G_0$ is the critical density of the universe, with $H_0$ being the Hubble constant and $G_0$ being the Newtonian gravitational constant, measured on Earth. $\rho_{\mathrm{GW}}$ is the GW energy density as a function of GW frequency $f$, defined as
\begin{equation}\label{pho_GR_simp}
\begin{aligned}
\rho_{\rm GW}=\frac{1}{16\pi G_0}\left\langle \dot{h}^{2}_{+}+\dot{h}^{2}_\times\right\rangle_{\lambda},
\end{aligned}
\end{equation}
with $\left\langle\cdots\right\rangle_{\lambda}$ means averaging over several GW wavelength. The total SGWB spectrum results from the cumulative contributions of all GW sources in the Universe, that cannot be detected as individual GW events by detectors. Such that, considering the source distribution and cosmological evolution, Eq.\,(\ref{omg_def}) can be expressed as a superposition of the energy densities radiated at each redshift \cite{Phinney:2001di,Nishizawa:2011eq},
\begin{equation}
\label{Omega_GW_population}
\begin{aligned}
\Omega_{\mathrm{GW}}(f)=\frac{f}{\rho_{c} H_0}\int_0^{\infty} d z \frac{\mathcal{R}_{\mathrm{GW}}(z)}{(1+z) \sqrt{\Omega_{m}(1+z)^3+\Omega_{\Lambda}}}\left\langle\frac{dE}{df}\right\rangle_{s},
\end{aligned}
\end{equation}
where $\mathcal{R}_{\mathrm{GW}}(z)$ is the merger rate of GW sources measured in the source frame. 
And, here $dE/df$ is the energy spectrum contributed by each astrophysical source, which is computed from the frequency-domain waveform, via
\begin{equation}\label{de/df_GR}
\left(\frac{dE}{df}\right)_{\rm GR} 
=\frac{\pi f^2}{2G_0}R^2
\left\langle\left|\tilde{h}_{+}(f)\right|^2+\left|\tilde{h}_{\times}(f)\right|^2\right\rangle_{\Omega},
\end{equation}
where $\left\langle\cdots\right\rangle_{\Omega}$ denotes the integration over the whole solid angle. $\tilde{h}_{+}(f)$ and $\tilde{h}_{\times}(f)$ are the GW polarization in the frequency domain, predicted by GR. $R$ is the distance between the GW source and the observer.
The $\langle\cdots\rangle_s$ in Eq.\,(\ref{Omega_GW_population}) denotes the averaged quantity over the population properties of compact binary systems, i.e., the chirp mass.

\begin{figure}[htbp]
\centering
\includegraphics[width=0.8\textwidth]{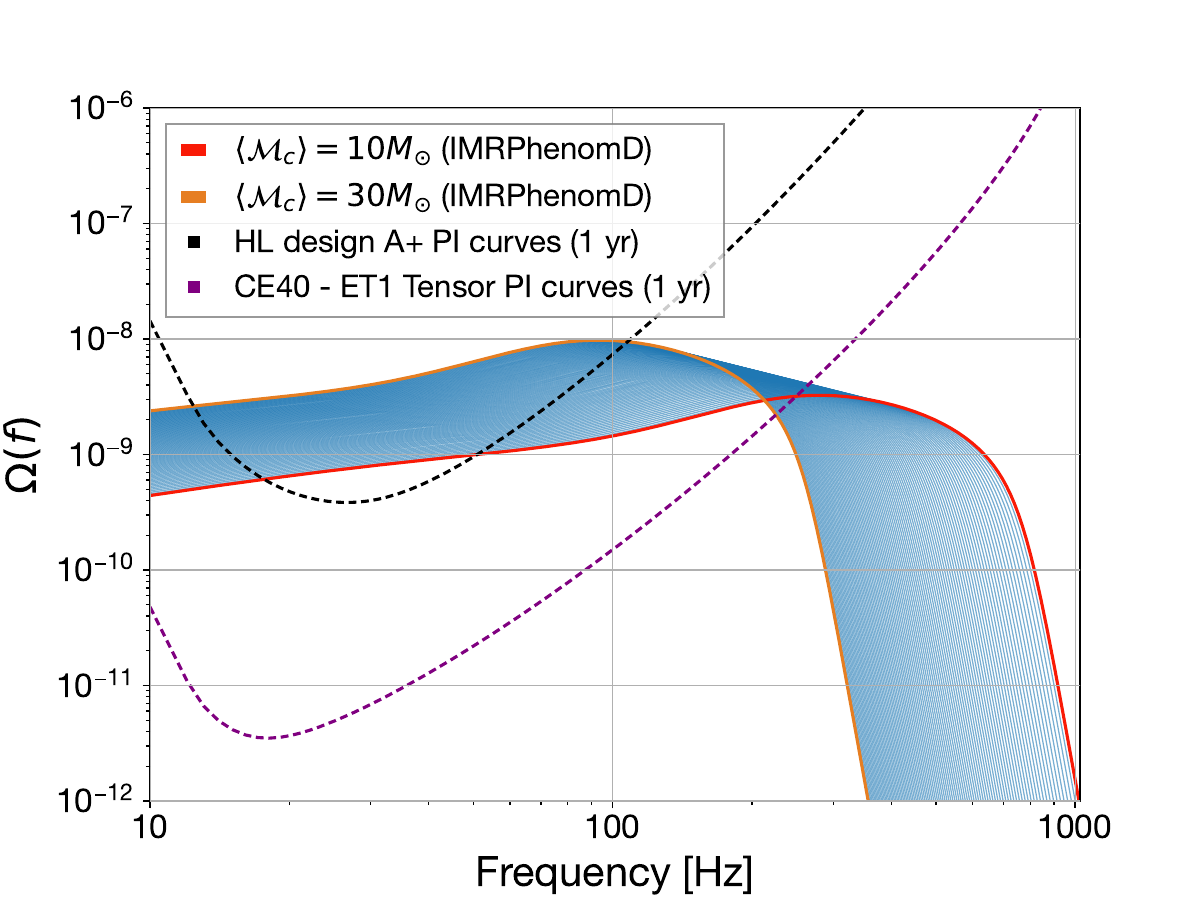}
\caption{
The predicted background spectra constructed using IMRPhenomD for BBH with average chirp mass range of $[10,30]M_{\odot}$ in GR framework. 
The designed one-year PI curve with the respond of tensor mode for the A+ sensitivity is shown as a dashed black line, while the sensitivity of the network consisting of the CE and ET detectors is depicted as a dashed purple line. 
The background spectra above these sensitivity curves are generally expected to have a SNR greater than unity. 
The merger and ringdown phases lead to a rapid decay in the background spectra, contributing a smaller stochastic background signal compared to the inspiral phase.}
\label{IMR_omegagw}
\end{figure}

To evaluate the detectability of the SGWB by the second and third-generation ground-based GW detectors, we first generate the frequency-domain waveforms through IMRPhenomD \cite{Husa:2015iqa,Khan:2015jqa}, which contain the entire GW signal from the inspiral, merger, and ringdown phases of BBH mergers.
The SGWB spectrum, $\Omega_{\rm GW}$, is computed using Eq.\,(\ref{Omega_GW_population}) and is depicted in Fig.\ref{IMR_omegagw}, assuming a mean chirp mass of $\left\langle\mathcal{M}_c\right\rangle$ within the range of $[10.0,30.0]M_{\odot}$.
Additionally, the merger rate of BBH systems are modeled based on the framework and parameters outlined in Section~\ref{sec:Population}, where the population models for different sources are completely discussed.
The power-law integrated (PI) sensitivity curves~\cite{Thrane:2013oya} for the aLIGO detectors at Livingston and Hanford, as well as for the Cosmic Explorer (CE) and Einstein Telescope (ET) detectors, are presented. 
By definition, the energy density spectrum, $\Omega_{\mathrm{GW}}$, lying above the PI curve corresponds to an expected signal-to-noise ratio SNR$>$1.

As illustrated in Fig.\ref{IMR_omegagw}, the SGWB generates its strongest signal within the frequency range of 10–200 Hz, corresponding to the inspiral phase of gravitational waves. This range is determined by considering the frequency at the innermost stable circular orbit (ISCO) as the endpoint of the inspiral phase for a binary merger~\cite{Maggiore:2007ulw}. For a binary black hole (BBH) with a total mass of
$m = 20M_{\odot}$, the ISCO frequency is approximately $(f_r)_{\rm ISCO} \simeq 200$ Hz.
Accordingly, our analysis focuses on the inspiral phase, as it provides the dominant contribution to the SGWB.

\subsection{SGWB spectrum in BD gravity}
\label{sec:SGWB spectrum in BD gravity}
This subsection investigates how the BD gravity modifies the SGWB spectrum. Under the BD framework, the gravity is described by the metric and scalar fields, denoted by $g_{\mu\nu}$ and $\phi$, respectively, simultaneously. The action of the massless BD theory in the Jordan frame is
\begin{equation}
\begin{aligned}\label{BD_action}
S=\frac{1}{16 \pi } \int d^4 x \sqrt{-g}\left[\phi(R-2\Lambda)-\frac{\omega_{\rm BD}}{\phi} g^{\mu \nu}\left(\partial_\mu \phi\right)\left(\partial_\nu \phi\right)\right]
+S_{m}\left[g_{\mu \nu}, \Psi_m\right],
\end{aligned}
\end{equation}
where $\omega_{\rm BD}$ denotes the scalar-tensor coupling, and $\Lambda$ is the cosmological constant. 
GR is recovered from BD theory in the limit $\omega_{\rm BD}\rightarrow+\infty$. $S_m$ represents the action of the matter, which does not depend on the scalar field $\phi$. According to the motivation of BD theory, the Newtonian gravitational constant $G_0$ is generalized to a scalar field $G$ that varies over spacetime, defined as
\begin{equation}
G\equiv\frac{1}{\phi_0},
\end{equation}
with $\phi_0$ is the vacuum expectation value of the scalar field $\phi$. 
Due to the cosmic expansion, $\phi_0$ is not constant but varies with the redshift $z$. By solving the modified Friedmann equation, one finds the evolution history of $\phi_0(z)$, shown as 
\begin{equation}
\label{phi-z}
\phi_0(z) = \phi_0(z=0)\left(\frac{1}{1+z} \right)^{\frac{1}{1+\omega_{\rm BD}}},
\end{equation}
with $\phi_0(z=0)$ being the present value of the background scalar field, $\phi_0(z=0)=1/G_0$. The GWs are the metric and scalar perturbation, denoted by $h_{\mu\nu}$ and $\delta\phi$, around the FLRW metric and $\phi_0$. BD theory modifies the SGWB spectrum in both GW generation and propagation. 

We first consider the binary merger on a local flat background. The binary consists of two inspiralling compact objects, characterized by two intrinsic parameters, the bare masses $m_{A}$ and sensitivities $s_{A}$ ($A=1,2$). The sensitivities depend on the internal structure of the compact objects. For white dwarfs, $s_A \simeq 0$, for neutron stars, $s_A\approx 0.2$, and for black holes, $s_A=0.5$.
We describe the gravitational waveforms perturbatively under the small coupling limit
\begin{equation}
\xi\equiv(3+2\omega_{\rm BD})^{-1}\ll1.
\end{equation}
In the source frame, the frequency-domain waveforms of plus and cross polarizations are calculated as
\cite{XingZhang:2017}
\begin{equation}
\label{frequency-domain-waveforms-plus-cross}
\begin{aligned}
\tilde{h}_{+}(f)
&=\sqrt{\frac{5\pi}{24}}
\frac{(G\mathcal{M}_c)^2}{R}u^{-7/6}
\left\{1+\xi\left[-\frac{1}{12}\Gamma^2+\frac{1}{3}\Delta-\frac{5}{48}\eta^{2/5}\mathcal{S}^2u^{-2/3}\right]\right\}\frac{1+\cos^2\iota}{2}e^{i\Psi_{+}},\\
\tilde{h}_{\times}(f)
&=\sqrt{\frac{5\pi}{24}}
\frac{(G\mathcal{M}_c)^2}{R}u^{-7/6}
\left\{1+\xi\left[-\frac{1}{12}\Gamma^2+\frac{1}{3}\Delta
-\frac{5}{48}\eta^{2/5}\mathcal{S}^2u^{-2/3}\right]\right\}\cos\iota
e^{i(\Psi_{+}+\pi/2)},
\end{aligned}
\end{equation}
up to the linear order of coupling $\xi$. The phase factors are 
\begin{equation}
\label{phase-plus-cross}
\Psi_{+}=2\pi f(t_c+R)+\frac{3}{128}u^{-5/3}
\left\{1-\xi\left[\frac{1}{6}\Gamma^2+\frac{2}{3}\Delta
+\frac{5}{42}\eta^{2/5}\mathcal{S}^2u^{-2/3}
\right]\right\}
-2\Phi_0-\frac{\pi}{4},
\end{equation}
with $\Phi_0$ being the initial phase. The dimensionless frequency is $u=\pi G\mathcal{M}_cf$. In Eqs.\,(\ref{frequency-domain-waveforms-plus-cross}) and (\ref{phase-plus-cross}), we have defined the symmetric mass ratio as $\eta\equiv m_1m_2/m^2$, $\iota$ is the angle between the line of sight and binary orbital plane, and $t_r\equiv t-R$ is the retarded time. Additionally, another three parameters depending on the sensitivities are defined by
\begin{equation}
\Delta\equiv(1-2s_1)(1-2s_2),\quad
\Gamma\equiv1-2(m_{2}s_{1}+m_{1}s_{2})/m,\quad\text{and}\quad
\mathcal{S}\equiv s_{1}-s_{2}.
\end{equation}
These results return to the GR case when coupling $\xi$ is zero.
Beyond plus and cross mode, the extra scalar field brings a new polarization, breathing mode, into the detected GW signals, whose radiated waveform can be expressed in a similar form,
\begin{equation}
\label{frequency-domain-waveforms-breathing}
\tilde{h}_{\rm b}(f)
=\sqrt{\frac{5\pi}{24}}
\frac{(G\mathcal{M}_c)^2}{R}u^{-7/6}
\cdot\frac{\xi}{2}
\left[-\eta^{1/5}\mathcal{S}u^{1/3}
\sin\iota e^{i\Psi^{(1)}_{\rm b}}
+\Gamma\sin^2\iota e^{i\Psi_{+}}\right],
\end{equation}
with phase factor being
\begin{equation}
\label{phase-breathing}
\Psi_{\rm b}^{(1)}=2\pi f(t_c+R)+\frac{3}{128}(2u)^{-5/3}
\left\{1-\xi\left[\frac{1}{6}\Gamma^2+\frac{2}{3}\Delta
+\frac{5}{42}\eta^{2/5}\mathcal{S}^2(2u)^{-2/3}
\right]\right\}-\Phi_0-\frac{\pi}{4}.
\end{equation}

We complete the discussion on GW generation and turn to the propagation process. Since the typical GW wavelength is much shorter than the cosmological scale, one can apply the geometric-optics approximation to construct the modified graviton number conservation law. By matching such conserved quantity at the overlap region of the cosmological background and source frame, one gets the modified waveforms in an expanding Universe. The readers can find more details in Ref.\,\cite{TanLiu:2023}. The main modifications can be added into frequency-domain waveforms (\ref{frequency-domain-waveforms-plus-cross}-\ref{phase-breathing}) in two steps. The first one is to replace the chirp mass and coalescence time by the redshifted one, i.e., $\mathcal{M}_{c}\rightarrow \mathcal{M}_{z}\equiv \mathcal{M}_{c}(1+z)$, and $t_c\rightarrow t_z$. The second, but more important, is to revise the luminosity distance $R$ into the modified one, defined by
\begin{equation}
D=R(1+z)\sqrt{\frac{\phi_0(z)}{\phi_0(z=0)}},
\end{equation}
rather than the standard form in the GR case.

Because of the BD modification, the GW energy density and energy spectrum in GR, shown in Eqs.\,(\ref{pho_GR_simp}) and (\ref{de/df_GR}), are revised as~\cite{Saffer:2017ywl}
\begin{equation}
\label{pho_BD_simp}
\rho_{\rm GW}=\frac{\phi_{0}}{16 \pi}\left[\left\langle \dot{h}^{2}_{+}+\dot{h}^{2}_\times\right\rangle_{\lambda} + \left(3+2\omega_{\rm BD}\right)\left\langle \dot{h}^{2}_{\rm b}\right\rangle_{\lambda}\right],
\end{equation}
and
\begin{equation}\label{de/df_BD}
\frac{dE}{df} = \frac{\pi f^2}{2}\phi_0 R^2
\left\langle\left|\tilde{h}_{+}(f)\right|^2+\left|\tilde{h}_{\times}(f)\right|^2+(3+2\omega_{\rm BD})
\left|\tilde{h}_{\rm b}(f)\right|^2\right\rangle_{\Omega}
\end{equation}
respectively. 

In GR, the energy spectrum contributed by the inspiraling binaries is given by~\cite{Phinney:2001di}
\begin{equation}
\label{de/df_gr}
\left(\frac{dE}{df}\right)_{\rm GR}=\frac{(\pi G_0)^{2/3}\mathcal{M}_{c}^{5/3}}{3(1+z)^{1/3}}f^{-1/3},
\end{equation}
where $f$ is the GW frequency observed on Earth today. 
The chirp mass is $\mathcal{M}_c\equiv\frac{\left(m_1 m_2\right)^{3/5}}{m^{1/5}}$, where $m_A$ are the individual masses of compact objects, and $m\equiv m_1+m_2$ is their total mass.
Unlike GR, the energy spectrum is contributed by both tensor and scalar polarizations, which are computed from the frequency-domain waveforms, listed in Eqs.\,(\ref{frequency-domain-waveforms-plus-cross} - \ref{phase-breathing}) with replacement $\mathcal{M}_{c}\rightarrow\mathcal{M}_{z}$, $t_c\rightarrow t_z$, and $R\rightarrow D$. The final result of $dE/df$ in BD gravity is
\begin{equation}
\frac{dE}{df}
=\left(\frac{dE}{df}\right)_{\rm Tensor}
+\left(\frac{dE}{df}\right)_{\rm Scalar}.
\end{equation}
where the tensor and scalar-sector contributions are
\begin{equation}
\label{dedf-tensor}
\left(\frac{dE}{df}\right)_{\rm Tensor}=(1+z)^{\frac{8}{3+3\omega_{\rm BD}}}\left\{1+\xi\left[\left(\frac{2}{3}\Delta
-\frac{1}{6}\Gamma^2\right)
-\frac{5}{24}
\eta^{2/5}\mathcal{S}^2
u^{-2/3}\right]
\right\}\left(\frac{dE}{df}\right)_{\rm GR},
\end{equation}
and
\begin{equation}
\label{dedf-scalar}
\begin{aligned}
\left(\frac{dE}{df}\right)_{\rm Scalar}&=\xi\cdot(1+z)^{\frac{8}{3+3\omega_{\rm BD}}}\Bigg\{
\frac{\Gamma^2}{6}+\frac{5}{24}\eta^{2/5}\mathcal{S}^2u^{-2/3}\\
&\qquad-\frac{15}{128}\pi\Gamma\eta^{1/5}\mathcal{S}u^{-1/3}
\cos\left[\Psi_{+}-\Psi_{b}^{(1)}\right]
\Bigg\}\left(\frac{dE}{df}\right)_{\rm GR},
\end{aligned}
\end{equation}
respectively. $(dE/df)_{\rm GR}$ has been shown in Eq.\,(\ref{de/df_GR}). The overall factor $(1+z)^{8/(3+3\omega_{\rm BD})}$ is sourced by the modified cosmic expansion. The scalar sector is proportional to the coupling $\xi$, representing a new degree of freedom due to the extra scalar field.

\begin{figure}[t]
\centering
\includegraphics[width=1.0\textwidth]{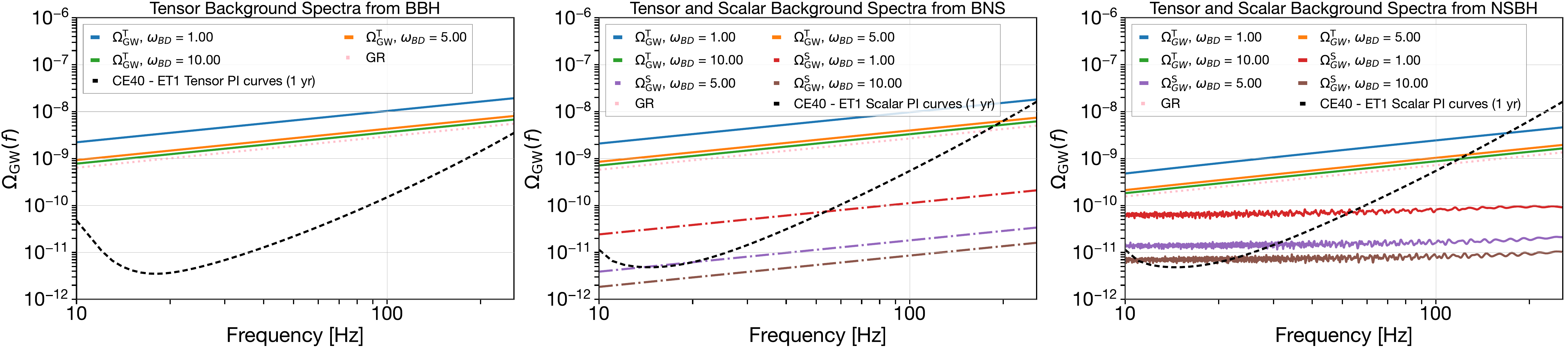}
\caption{
We present the background spectra contribution from  BBH, BNS, and NSBH mergers, utilizing the frequency-domain inspiral waveforms within the BD theory as described in the text and the population models described in Section~\ref{sec:Population}. The tensor and scalar mode backgrounds, corresponding to typical values of $\omega_{\rm BD} = [1, 5, 10]$, are represented by solid and dash-dotted lines, respectively. For comparison, the background in GR is shown as a dotted line (pink).
The prospective one-year PI curves for the sensitivity of the CE and ET networks, shown as black dashed lines, represent tensor mode responses in BBH mergers and scalar mode responses in BNS and NSBH mergers, indicating the potential detection of the corresponding polarization modes with SNR $>1$ when the background spectra lie above these lines.
}
\label{BBH_BNS_NSBH_omegagw}
\end{figure}

In this work, we consider all three usual classes of binary systems, the BBH, BNS, and NSBH. For the first class, two individual black holes have equal sensitivities, i.e., $s_1\approx s_2=0.5$, and then we have $\mathcal{S}=\Delta=\Gamma=0$. This means there are no BD modifications and scalar polarization in GW generation. Therefore, the energy spectrum is revised through GW propagation, which can be written as
\begin{equation}
\left(\frac{dE}{df}\right)_{\rm BBH,Tensor}
=(1+z)^{\frac{8}{3+3\omega_{\rm BD}}}\left(\frac{dE}{df}\right)_{\rm GR}.
\end{equation}
For the second class, we assume that all of the neutron stars have the same sensitivity~\cite{Will:1989sk,Sampson:2014qqa}, i.e., $s_1\approx s_2=0.2$, resulting in $\mathcal{S}=0$ and $\Delta,\Gamma\neq0$, and then the energy spectrum is 
\begin{equation}
\begin{aligned}
\left(\frac{dE}{df}\right)_{\rm BNS,Tensor}&=(1+z)^{\frac{8}{3+3\omega_{\rm BD}}}\left(1+\frac{9}{50}\xi\right)\left(\frac{dE}{df}\right)_{\rm GR},\\
\text{and}\quad\left(\frac{dE}{df}\right)_{\rm BNS,Scalar}&=(1+z)^{\frac{8}{3+3\omega_{\rm BD}}}\left(\frac{3}{50}\xi\right)\left(\frac{dE}{df}\right)_{\rm GR}.
\end{aligned}
\end{equation}
The above two sectors only change the amplitude of the SGWB spectrum, without altering the spectral index. 
It is evident that the primary contribution of the correction arises from the propagation phase rather than the generation phase.
The third class is more complicated. Due to the different sensitivities of objects, we have $\Delta=0$ but $\Gamma,\mathcal{S}\neq0$. In addition to correcting the SGWB amplitude and exciting scalar modes, the spectral index will also change in this sector, and a frequency-dependent oscillation [shown in $\cos(\cdots)$ in Eq.\,(\ref{dedf-scalar})] will be generated due to the phase difference between plus and breathing modes.

The background spectra for such three classes binaries is shown in Fig.\ref{BBH_BNS_NSBH_omegagw}. 
In the tensor mode sector, BBH systems are expected to dominate the contribution to the background spectra or, at least, produce an intensity comparable to that of BNS systems. 
In contrast, the overall contribution from NSBH systems remains negligible, accounting for only approximately 10\% of the total tensor mode background, consistent with previous studies~\cite{kowalska2015effect, LIGOScientific:2016hpm, LIGOScientific:2017zlf, Li:2023awp}.
Furthermore, although NSBH systems modify the spectral index, these changes are not significantly evident within the tensor sector.
In the scalar mode sector, as previously discussed, there is no scalar radiation in BBH system, in contrast to BNS and NSBH systems. 
NSBH systems are expected to contribute a scalar background comparable to, or slightly larger than, that of BNS systems. 
Notably, NSBH systems could induce significant modifications to the spectral index of the scalar gravitational-wave background and introduce oscillatory behavior.
Detecting these phenomena may provide valuable insights for identifying potential deviations from GR.

\subsection{Stochastic signals}
\begin{figure}[t]
\centering
\includegraphics[width=1\textwidth]{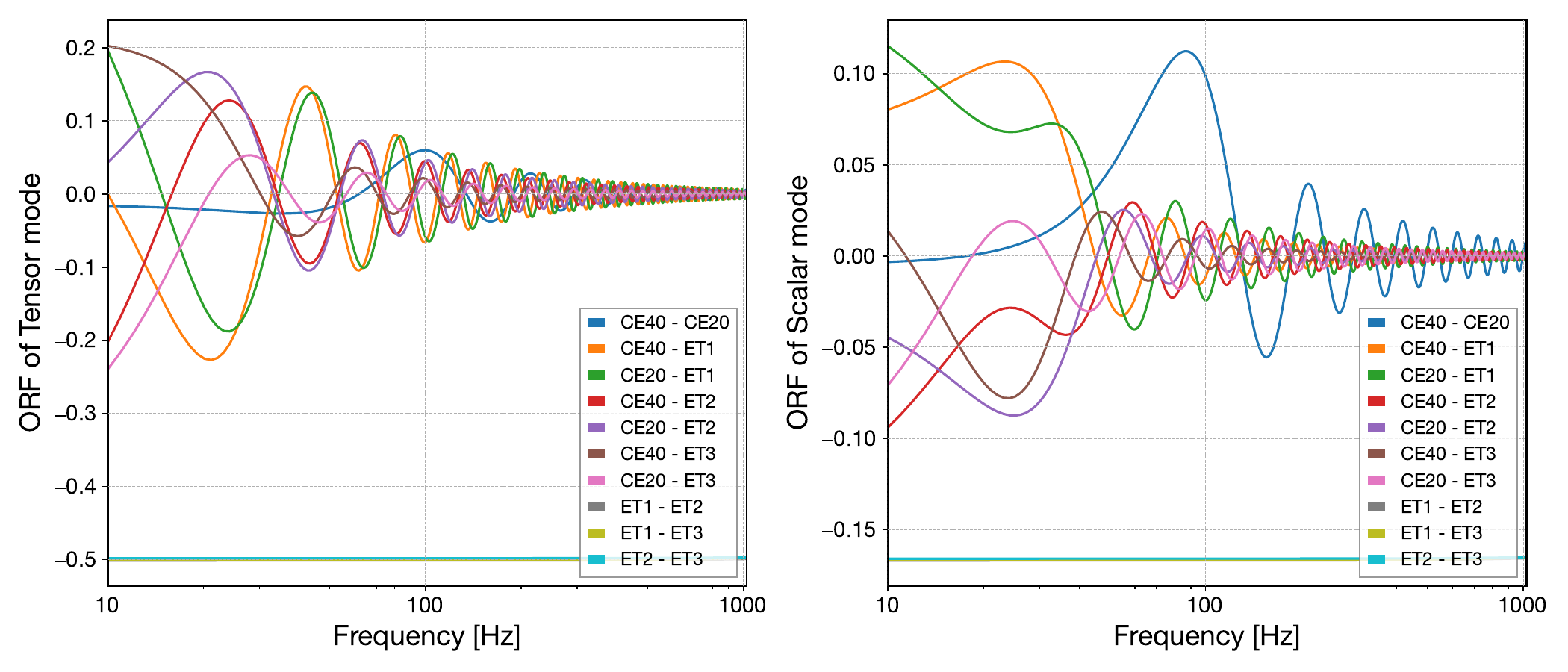}
\caption{The overlap reduction functions for the tensor mode (left) and the scalar mode (right) in 2CE and ET baselines indicate higher sensitivity to the stochastic background. 
Apart from the constant overlap reduction function for the three independent interferometers of the ET detector, the overlap reduction functions oscillate around zero, resulting in low sensitivity in such frequency range.
\label{fig_orf}}
\end{figure}

GW detectors do not respond to the SGWB spectrum directly, instead, they measure the GW amplitude at each instrument. Consequently, we need a theory-dependent mapping that relates GW amplitudes to the SGWB spectrum. 
The metric perturbation can be expressed as a linear superposition of plane gravitational waves, allowing for a decomposition of the signal into frequency $f$, propagation direction $\hat{\mathbf{n}}$ and polarization $A$~\cite{Maggiore:2007ulw,Nishizawa:2009bf}:
\begin{equation}\label{pwe}
\begin{aligned}
h_{ij}(t, \mathbf{x})=\sum_{A=+,\times,{b}} \int_{-\infty}^{\infty} d f \int d^2 \hat{\mathbf{n}} \tilde{h}_A(f, \hat{\mathbf{n}}) e_{ij}^A(\hat{\mathbf{n}}) e^{2 \pi i f(t-\hat{\mathbf{n}} \cdot \mathbf{x} / c)},
\end{aligned}
\end{equation}
where $e_{ij}^A$ is the polarization tensors, especially, $A=+,\times,\mathrm{b}$ in this work. 
Assuming that the SGWB is stationary, Gaussian, and isotropic, with uncorrelated polarizations, the second moment of the stochastic signals can be directly expressed in terms of the power spectral density $S_{A}(f)$ as
\begin{equation}\label{topsd}
\begin{aligned}
\left\langle \tilde{h}_A^*(f, \hat{\mathbf{n}}) \tilde{h}_{A^{\prime}}\left(f', \hat{\mathbf{n}}^{\prime}\right)\right\rangle=\delta(f-f^{\prime})\frac{1}{4\pi}\delta^2\left(\hat{\mathbf{n}}-\hat{\mathbf{n}}'\right) \delta_{AA'} \frac{1}{2} S_A(f),
\end{aligned}
\end{equation}
where the factor of $1/2$ indicates that this equation defines the one-sided power spectral density for each polarization mode.
Using Eqs.\,(\ref{omg_def}) and (\ref{pho_BD_simp}), one can relate SGWB spectrum in plus, cross, and breathing polarizations to their strain power-spectral densities via~\cite{Isi:2018miq}
\begin{equation}
\label{rela_ed_psd}
\Omega_A(f) = \frac{2 \pi^2f^3}{3 H_0^2}\lambda_A(f) S_A(f), \quad
\lambda_A= \begin{cases}\xi^{-1} & \text { if } A=\mathrm{b}, \\ 1 & \text { if } A=+, \times,\end{cases}
\end{equation}
The total spectrum is the sum of the contribution by three polarizations, $\Omega_{\rm GW}=\Omega_{+}+\Omega_{\times}+\Omega_{\rm b}$. 
The response of the detector $I$ to a passing GW signal is written as the linear combination of each polarization mode~\cite{Nishizawa:2009bf}, 
\begin{equation}
\begin{aligned}
\tilde{h}_I(f)=\int d^2 \hat{\mathbf{n}} \sum_A F_I^A(\hat{\mathbf{n}}) \tilde{h}_A(f, \hat{\mathbf{n}}) e^{-2 \pi i f \hat{\mathbf{n}} \cdot \mathbf{x}_I},
\end{aligned}
\end{equation}
where $F_I^A(\hat{n})$ is the antenna pattern functions of detector $I$ for polarization $A$. In the Fourier domain, the cross-correlation between $\tilde{h}_I(f)$ and $\tilde{h}_J(f)$ of detectors $I$,$J$ can be expressed in terms of the second moment of stochastic signals,
\begin{equation}
\begin{aligned}
\label{hIhJ}
&\left\langle\tilde{h}_I^*(f) \tilde{h}_{J}\left(f^{\prime}\right)\right\rangle =  \int  d^2 \hat{\mathbf{n}} ~d^2 \hat{\mathbf{n}}^{\prime} \sum_{A A^{\prime}}\left\langle\tilde{h}_A^*(f, \hat{\mathbf{n}}) \tilde{h}_{A^{\prime}}\left(f^{\prime}, \hat{\mathbf{n}}^{\prime}\right)\right\rangle \\
&\qquad\qquad\qquad\qquad\times F_I^{* A}(\hat{\mathbf{n}}) F_{J}^{A^{\prime}}\left(\hat{\mathbf{n}}^{\prime}\right) e^{-2\pi i( f^{\prime} \hat{\mathbf{n}}^{\prime}\cdot  \mathbf{x}_J/ c-f\hat{\mathbf{n}} \cdot \mathbf{x}_I / c)}.
\end{aligned}
\end{equation}
Inserting Eq.\,(\ref{topsd}) into (\ref{hIhJ}), and using (\ref{rela_ed_psd}), we would write it as
\begin{equation}
\begin{aligned}
&\left\langle\tilde{h}_I^*(f) \tilde{h}_{J}\left(f^{\prime}\right)\right\rangle= \frac{3 H_0^2}{20 \pi^2} f^{-3}\delta\left(f-f^{\prime}\right)\widehat{\Omega}_{\rm GW},
\end{aligned}
\end{equation}
Here, the canonical SGWB spectrum, $\widehat{\Omega}_{\rm GW}$, composed of the energy densities from the tensor part $\Omega_{\mathrm{GW}}^\mathrm{T}\equiv\Omega_{\mathrm{GW}}^{+}+\Omega_{\mathrm{GW}}^{\times}$ (with assumption $\Omega_{\mathrm{GW}}^{+}=\Omega_{\mathrm{GW}}^{\times}$~\cite{Nishizawa:2009bf}) and the scalar part $\Omega_{\mathrm{GW}}^\mathrm{S}\equiv\Omega_{\mathrm{GW}}^{\rm b}$, can be expressed as
\begin{equation}
\label{widehat_Omega}
\widehat{\Omega}_{\rm GW}\equiv
\Omega_{\mathrm{GW}}^\mathrm{T} \gamma_{IJ}^\mathrm{T}+\xi\cdot\Omega_{\mathrm{GW}}^\mathrm{S} \gamma_{IJ}^\mathrm{S},
\end{equation}
where we have defined the normalized overlap reduction functions (ORF) for the tensor component, $\gamma_{IJ}^\mathrm{T}(f)$, and the scalar component with a absorbed 1/3 factor, $\gamma_{IJ}^\mathrm{S}(f)$, as described in~\cite{Nishizawa:2009bf, Allen:1997ad}, 
\begin{equation}
\begin{aligned}
&\gamma_{I J}^{\rm T}(f) \equiv \frac{5}{2} \int \frac{d^2 \hat{\mathbf{n}}}{4 \pi} \left(F_I^{+} F_J^{+}+F_I^{\times} F_J^{\times}\right) e^{2 \pi i f  \hat{\mathbf{n}} \cdot (\mathbf{x}_I-\mathbf{x}_J)},\\
&\gamma_{I J}^{\rm S}(f) \equiv 5 \int \frac{d^2 \hat{\mathbf{n}}}{4 \pi} \left(F_I^{\rm b} F_J^{\rm b}\right) e^{2 \pi i f  \hat{\mathbf{n}} \cdot (\mathbf{x}_I-\mathbf{x}_J)}.
\end{aligned}
\end{equation}
The overlap reduction functions characterize the capability of detectors to retain information about GW signals with various polarization modes. 
In Fig.~\ref{fig_orf}, we present the overlap reduction functions for the CE and ET networks. Notably, the ET consists of three separate interferometers positioned at the vertices of an equilateral triangle. 
The overlap reduction function for the scalar mode shows a magnitude nearly comparable to that of the tensor mode, indicating similar detection capabilities for the two modes.

\subsection{Cross-correlation statistic}
The cross-correlation of detector pairs is a commonly employed method for detecting the SGWB. 
We denote $\tilde{s}_{I(J)}$ as the Fourier-transformed strain signal measured by detector $I(J)$, which includes contributions from both physical GW signals and instrumental noise, $\tilde{s}_{I}=\tilde{h}_{I}+\tilde{n}_{I}$. 
We assume the noise in each detector is independent and satisfies $\langle\tilde{h}_I\tilde{n}_J\rangle=\langle\tilde{n}_I\tilde{h}_J\rangle=\langle\tilde{n}_I\tilde{n}_J\rangle=0$. We separate the whole signal into segments, with frequency bin $\delta f$ and duration $T$ in each segment. For the $i$-th segment in $[f,f+\delta f]$, an unbiased and minimal-variance optimal estimator for $\Omega_{\mathrm{GW}}$ is given by~\cite{LIGOScientific:2014sej,Romano:2016dpx,LIGOScientific:2018czr,Callister:2017ocg,Renzini:2023qtj}
\begin{equation}
\begin{aligned}
\hat{C}_{IJ(i)}(f)=\frac{2}{T}\frac{10 \pi^2}{3 H_0^2}f^3 \operatorname{Re}\left[\tilde{s}_{I}^*(f) \tilde{s}_{J}(f)\right].
\end{aligned}
\end{equation}
Averaging over an ensemble of noise realizations, the expectation value of this estimator is just the canonical SGWB spectrum,
\begin{equation}
\begin{aligned}
\label{estimator_expectation_value}
\left\langle\hat{C}_{IJ(i)}(f)\right\rangle
=\widehat{\Omega}_{\rm GW}.
\end{aligned}
\end{equation}
with the corresponding variance~\cite{LIGOScientific:2018czr}:
\begin{equation}
\begin{aligned}
\sigma_{IJ(i)}^2(f)&=\frac{1}{2T\delta f}\left(\frac{10 \pi^2}{3 H_0^2}\right)^2 f^6 P_I(f) P_J(f).
\end{aligned}
\end{equation}
Here, $P_{I(J)}(f)$ is the one-sided auto-power spectral density of detector $I(J)$, defined by
\begin{equation}
\left\langle\tilde{s}_{I}^*(f) \tilde{s}_{I}\left(f^{\prime}\right)\right\rangle=\frac{1}{2} \delta\left(f-f^{\prime}\right) P_{I}(f).
\end{equation}
The final estimator and variance are optimally combined by the weighted sum from each segment via~\cite{LIGOScientific:2018czr}
\begin{equation}\label{final estimator}
\hat{C}_{IJ}(f)=\frac{\sum_i \hat{C}_{IJ(i)}(f) \sigma_{IJ(i)}^{-2}(f)}{\sum_i \sigma_{IJ(i)}^{-2}(f)}, \ \sigma_{IJ}^{-2}(f)=\sum_i \sigma_{IJ(i)}^{-2}(f),
\end{equation}
where $\hat{C}_{IJ(i)}$ and $\sigma_{IJ(i)}$ denote the individual segment estimators of the detector pairs $IJ$ and their inverse variances, respectively.
With the estimate of the SGWB spectrum $\hat{C}_{IJ}(f)$ and variance $\sigma_{IJ}^{2}(f)$, we perform parameter estimation and fit the model. 
The Gaussian likelihood is formed by combing the spectrum from each baseline $IJ$ as
\begin{equation}
p\left(\hat{C}_{IJ} \mid \boldsymbol{\Theta}\right) \propto \exp \left[-\frac{1}{2} \sum_{I J}\sum_f\left(\frac{\hat{C}_{IJ}(f)-\widehat{\Omega}_{\rm GW}(f, \boldsymbol{\Theta})}{\sigma_{IJ}(f)}\right)^2\right],
\end{equation}
where $\widehat{\Omega}_{\rm GW}(f, \boldsymbol{\Theta})$ contains the information of SGWB-spectrum model and $\boldsymbol{\Theta}$ are its parameters. Given the likelihood, one can estimate the posterior distribution of the parameters of the model using Bayes theorem,
\begin{equation}
p\left(\boldsymbol{\Theta}\mid \hat{C}_{IJ}\right) \propto p\left(\hat{C}_{IJ} \mid \boldsymbol{\Theta}\right)
p(\boldsymbol{\Theta}),
\end{equation}
where $p(\boldsymbol{\Theta})$ is the prior distribution on the parameters $\boldsymbol{\Theta}$.

\section{Simulation and Parameter Estimation}
\label{sec:Simulation and Parameter Estimate}

\subsection{Population Models}
\label{sec:Population}
Recalling the previous sections, the total gravitational wave energy density $\Omega_{\mathrm{GW}}$ depends on the merger rate $\mathcal{R}_{\mathrm{GW}}(z)$ of compact binary coalescences, as well as the energy spectrum $\langle\frac{dE}{df}\rangle_{s}$, which is the function of mass of compact binary.
 
The merger rate of binary systems is given by the following form~\cite{Madau:2014bja}: 
\begin{equation}
\begin{aligned}
\mathcal{R}_{\mathrm{GW}}(z)=\frac{\mathcal{R}_{0}}{\mathcal{C}} \frac{(1+z)^{\alpha_z}}{1+\left(\frac{1+z}{1+z_p}\right)^{\alpha_z+\beta_z}}.
\end{aligned}
\end{equation}
Here, $\mathcal{R}_{0}$ is the local rate of binary systems evaluated at redshift $z = 0$ and $\mathcal{C}$ is a normalization constant to ensure $\mathcal{R}_{\mathrm{GW}}(0) = \mathcal{R}_{0}$.
We adopt parameters
$\alpha_z=2.7$, $\beta_z=2.4$, $z_{p} = 2.0$, consistent with Refs~\cite{Madau:2014bja,KAGRA:2021duu}.
The local rates for BBH, BNS, and NSBH mergers are $\mathcal{R}^{\mathrm{BBH}}_{0}=31\ \mathrm{Gpc}^{-3} \mathrm{yr}^{-1}$, $\mathcal{R}^{\mathrm{BNS}}_{0}=855\ \mathrm{Gpc}^{-3} \mathrm{yr}^{-1}$, and $\mathcal{R}^{\mathrm{NSBH}}_{0}=74\ \mathrm{Gpc}^{-3} \mathrm{yr}^{-1}$, respectively~\cite{KAGRA:2021duu}.
We truncate the merger rate at $z_{max}=10$, assuming it to be zero at higher redshifts, as the contribution of such distant systems to the overall gravitational wave background is relatively small.

Note that Eq.(\ref{de/df_gr}) assumes an average chirp mass for all binaries, meaning it does not account for variations in the chirp mass across different binary systems. 
Given that the total gravitational wave energy density is calculated from the cumulative contributions of various sources, a more general form of Eq.(\ref{de/df_gr}) is to replace $\mathcal{M}_{c}^{5/3}$ with $\langle \mathcal{M}_{c}^{5/3} \rangle$, where $\langle \mathcal{M}_{c}^{5/3} \rangle$ is determined by the mass distribution of the binary systems.

For BBH binaries, we assume a mass distribution following the Power Law + Peak model~\cite{Talbot:2018cva}, where the average chirp mass is computed as:
\begin{equation}
    \langle \mathcal{M}_{c}^{5/3} \rangle_{\mathrm{BBH}} = \int \mathcal{M}_{c}^{5/3} p(m_1,m_2|{\bf {\mathrm \Lambda_{m}}}) d m_1 dm_2.
\end{equation}
Here, $p(m_1,m_2|{\bf {\mathrm \Lambda_{m}}})$ represents the mass function of this model, where
\begin{equation}
{\bf {\mathrm \Lambda_{m}}}=\left\{\alpha, \beta, m_{\min}, m_{\max}, \delta_{\rm m}, \lambda_{\rm peak},\mu_{\rm m},\sigma_{\rm m}\right\}
\end{equation}
denotes the hyper-parameters of the model.
To ensure consistency with the distribution described in Ref~\cite{KAGRA:2021duu}, we adopt the following parameters: 
$\alpha=3.5$, $\beta=1$, $m_{\min}=5M_{\odot}$, $m_{\max}=80M_{\odot}$, $\delta_{\mathrm m}=5M_{\odot}$, $\lambda_{\rm peak}=0.035$, $\mu_{\mathrm m}=34M_{\odot}$, and $\sigma_{\mathrm m}=5 M_{\odot}$.
We then obtain $\langle \mathcal{M}_{c}^{5/3} \rangle_{\mathrm{BBH}}= 60.8 M_{\odot}^{5/3}$, which will be treated as the only free parameter in our Bayesian analysis of the AGWB.

For the BNS binaries, we simply assume both component masses follow a uniform distribution in (1.2, 2.3) $M_{\odot}$ and are randomly paired, which is consistent with the currently available data \citep{Li:2021oyk,Madau:2014bja,Landry:2021hvl}. 
This yields $\langle \mathcal{M}_{c}^{5/3} \rangle_{\rm BNS} = 2.0 M_{\odot}^{5/3}$.
Since the sensitivities depend on the equation of state (EoS) and
the mass. 
For a fixed EoS, the sensitivities depend only on the mass of the object. 
Given the narrow mass range of neutron stars under consideration, it is reasonable to adopt fixed sensitivities of $s_1 = s_2=0.2$~\cite{Will:1989sk,Sampson:2014qqa}.

Different from BBH and BNS systems, the background spectrum for NSBH binaries cannot be calculated solely in terms of the characteristic parameter $\langle \mathcal{M}_{c}^{5/3} \rangle$.
NSBH binaries introduces modifications to the spectrum, notably generating oscillatory effects on the scalar part.
To streamline the analysis while retaining all essential effects, we consider the average masses of the binaries, omitting the mass distribution from the integral.
We adopt an average black hole mass of $\langle M_{\mathrm{BH}} \rangle = 8.36 M_{\odot}$ and an average neutron star mass of $\langle M_{\mathrm{NS}} \rangle = 1.60 M_{\odot}$ among NSBH mergers~\cite{Chen:2024xkv}.
The phase at coalescence, $\Phi_0$, is an integration constant, chosen to be zero.
Additionally, we set the initial sensitivities of black holes and neutron stars to $s_{\rm BH} = 0.5$ and $s_{\rm NS} = 0.2$, respectively.

\subsection{Simulations and spectral calculation}
\label{sec:Simulations and spectral calculation}
In this work, we aim to constrain BD theory through two approaches.
The first involves considering the amplitude modifications to the background spectrum caused by waveform corrections under BD gravity.
The second is to search the scalar mode, which is predicted by the BNS and NSBH systems in BD gravity. 
This approach is designed to explore the potential for detecting scalar gravitational waves with the network of future gravitational wave detectors. 
 
We generated two sets of one-year simulated AGWB data by \Pygwb~\cite{Renzini:2023qtj} for the two approaches. 
Based on GR, the first set adopts a fiducial universe and classical populations of BBHs which are consistent with gravitational-wave observations from LIGO-Virgo-KAGRA up to GWTC-3~\cite{KAGRA:2021duu}.
The second set of simulated data is generated using a power-law model, which is fitted to the tensor and scalar gravitational wave energy spectra derived from population models of the BBH, BNS, and NSBH mergers.

We employ the 2CE+ET detector network, where 2CE includes CE40 and CE20, referring to the proposed 40-km and 20-km arm configurations of the CE, respectively, while ET corresponds to the proposed triangular configuration, consisting of three nested detectors~\cite{LIGOsens}.
To minimize the computational cost of the simulations while maintaining optimal detection prospects, we configured the network as \{CE40, CE20, ET1\} for the 2CE and ET detectors.
The following procedure in each detector is applied to both simulated datasets as described below.

For the generation of background signals, a sampling frequency of 1024 Hz is employed. 
The simulated datasets for each interferometer are first constructed in the frequency domain and then transformed into the time domain using an inverse discrete Fourier transform. 
And we assume that the noise is uncorrelated between all detectors.

The total time series is divided into intervals, with each interval further segmented, to optimize memory usage and computational efficiency.
Specifically, we use 60 segments, each containing 256 seconds of data.

To obtain the power and cross spectra of the detectors for each interval, the segmented time series are Hann-windowed and overlapped by 50\% before calculating the Fourier transforms~\cite{Renzini:2023qtj}. 
The resulting spectra are then coarse-grained to a frequency resolution of 1/64 Hz and restricted to the 10-200 Hz frequency band, which contributes most significantly to the expected signal-to-noise ratio.
The calculated power and cross spectra for each interval are combined by Eq.(\ref{final estimator}) to obtain the optimal estimator $\hat{C}_{I J}(f)$ and its corresponding variance $\sigma_{I J}^2(f)$. 
Following the convention adopted in~\cite{Renzini:2023qtj}, we recover the overlap reduction functions in both the optimal estimator $\hat{C}_{I J}(f)$ and its associated variance $\sigma_{I J}^2(f)$. 
In this case, for scenarios where the total background consists of a single polarization mode (e.g., tensor or scalar), both quantities depend exclusively on the stochastic background components.

\begin{table*}[t]
    \centering
    \scriptsize
    \caption{Parameters and their prior distributions used in the analyses of Dataset \uppercase\expandafter{\romannumeral1}.}
    \label{tab:priors}
    \begin{tabular}{c p{4cm} c c}
        \hline\hline
        \textbf{Parameter} & \textbf{Description} & \textbf{Fiducial values} & \textbf{Prior} \\
        \hline
        \rule{0pt}{15pt}
        $\langle \mathcal{M}_{c}^{5/3} \rangle \  [M_{\odot}^{5/3}] $ & The average of the chirp mass raised to the power of 5/3. & 60.8 & Uniform $[1, 200]$ \\
        $\mathrm{Log}_{10}(1/\omega_{BD})$ & The logarithm of the reciprocal of the parameter in BD theory. &$-10$ & Uniform $[-10, 10]$ \\
        $\mathcal{R}^{\mathrm{BBH}}_{0} \  [\mathrm{Gpc}^{-3} \mathrm{yr}^{-1}]$ & The local BBH merger rate at $z=0$. & $31$ & Uniform $[1, 100]$ \\
        $\alpha_z$ & The leading slope of the BBH merger rate $\mathcal{R}^{\mathrm{BBH}}(z)$. &$2.7$ & Gaussian $(3, 1.5)$ \\
        $\beta_z$ & The trailing slope of the BBH merger rate $\mathcal{R}^{\mathrm{BBH}}(z)$. & $2.4$ & Gaussian $(1, 10)$ \\
        $z_{p}$ & The peak redshift of the BBH merger rate $\mathcal{R}^{\mathrm{BBH}}(z)$. &$2.0$ & \parbox[c]{4cm}{\centering Truncated Gaussian $(0.5, 4)$ \\ \centering bounded within (0, 30)} \\
        \hline\hline
    \end{tabular}
\end{table*}

\subsection{Waveform-Based Constraints}
In this subsection, we investigate the potential to constrain BD gravity through modifications to the background spectral amplitude arising from waveform corrections.
As discussed in Sec.~\ref{sec:SGWB spectrum in BD gravity}, the redshift-dependent amplification of the background spectrum primarily results from propagation effects in BBH and BNS mergers. 
These effects leave the spectral index of the background unchanged while introducing a degeneracy between the BD gravity parameter and the population properties, including the mass distribution and merger rate of binary systems.
However, the detectability of the background spectrum is primarily determined by its intensity relative to the detector noise and 
the spectra with higher intensities can overshadow weaker ones.
Therefore, while NSBH mergers could introduce significant modifications to the spectral index of the scalar background and generate oscillatory behavior, their contribution to the total background spectrum is significantly smaller compared to BBH and BNS systems. As a result, constraining BD gravity through background spectral corrections based on NSBH mergers remains challenging.

Based on the above considerations, we focus exclusively on BBH systems, as they are expected to provide the strongest contribution to the AGWB, making them the most detectable signal amidst noise.
Additionally, compared to BNS systems, the extensive observational data on BBH systems provides a more comprehensive understanding of their population properties, allowing us to evaluate the influence of these properties on constraining BD gravity.

\begin{figure}[htbp]
\centering
\includegraphics[width=.46\textwidth]{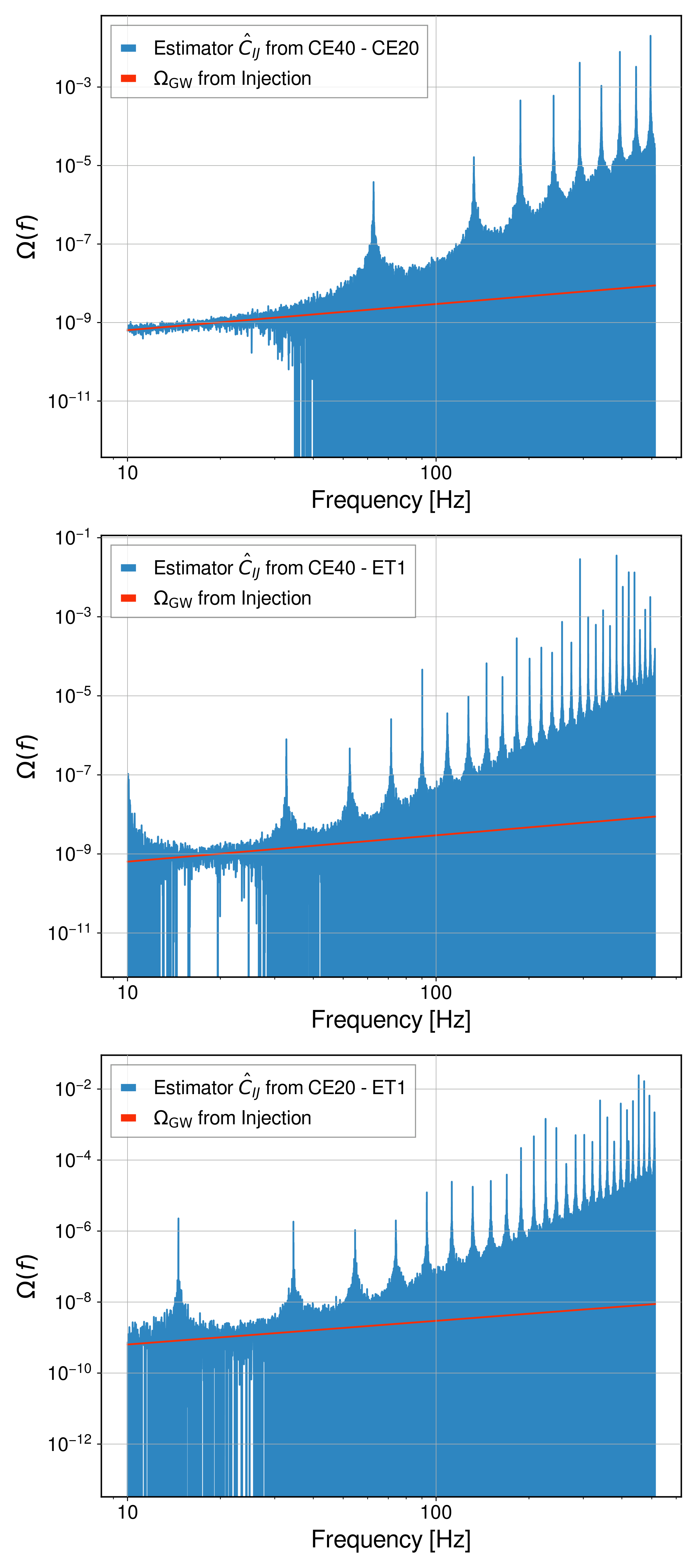}
\qquad
\includegraphics[width=.46\textwidth]{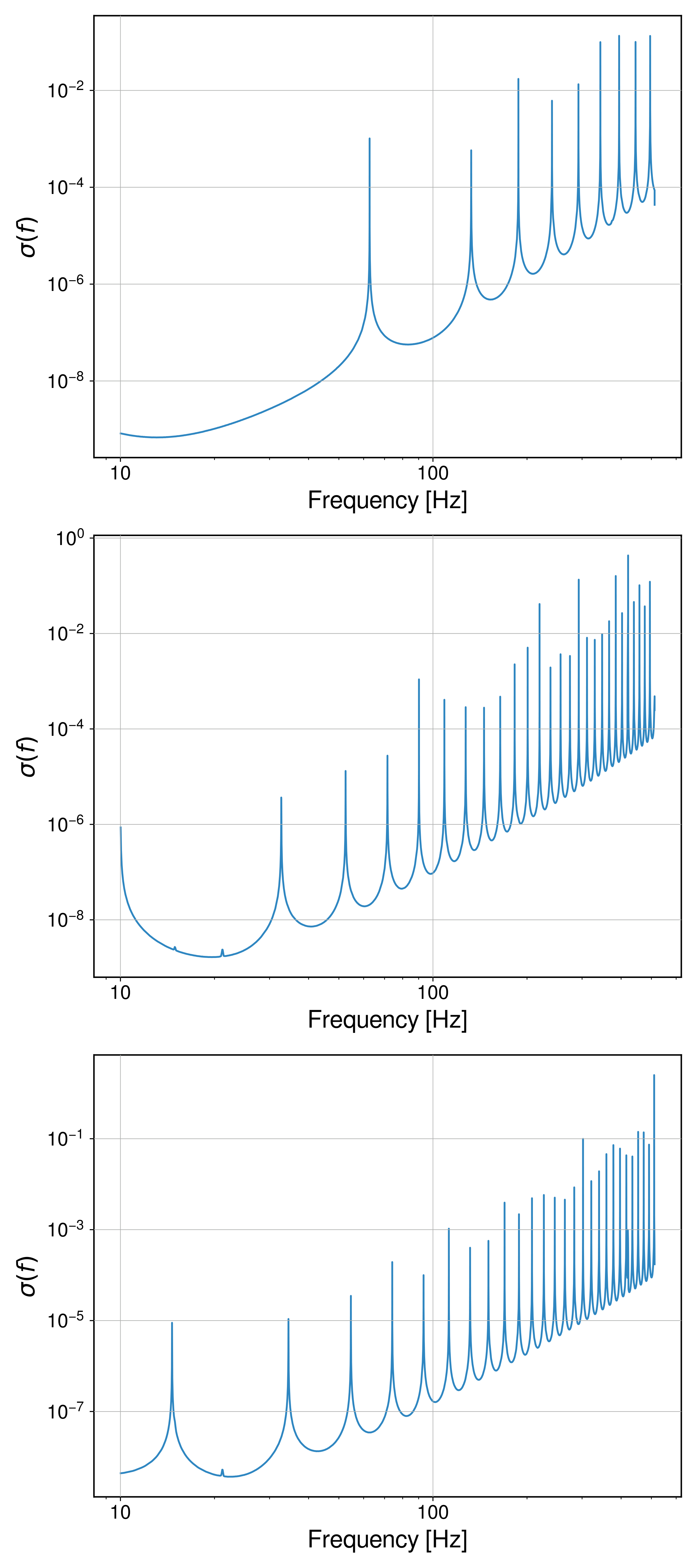}
\caption{The estimated background spectra, $\hat{C}_{IJ}(f)$, and their corresponding variances, $\sigma_{IJ}^{2}(f)$, derived from three sets of baselines composed of 2CE and ET, are presented in the left (blue) and right panels, respectively. 
These results are obtained from one-year mock datasets (Dataset \uppercase\expandafter{\romannumeral1}). The fiducial SGWB spectrum (red) is plotted in the left panel for comparison with the estimations.}
\label{fig:omega_estimator_injection}
\end{figure}

We first generate a one-year mock dataset (denoted by ``Dataset \uppercase\expandafter{\romannumeral1}"), consisting of BBH mergers solely, with the population model described in Sec \ref{sec:Population}. 
For each BBH system, the emitted GW signal is expressed in Eq.\,(\ref{dedf-tensor}) in BD theory. 
As described in Section~\ref{sec:Simulations and spectral calculation}, the total SGWB signal is injected into each detector, with the associated detector noise accounted for, to obtain the optimal estimator and corresponding variance.
Fig.~\ref{fig:omega_estimator_injection} presents the injected SGWB energy density $\Omega_{\mathrm{GW}}(f)$, calculated in GR using the fiducial population and cosmology model.
It also shows the corresponding estimator and variance from the baseline in the network of interferometers.
The injection of $\Omega_{\mathrm{GW}}(f)$ is well restored, particularly in the frequency band below 100 Hz.
Globally, the variance increases with frequency due to the rising noise PSD of the detector. Locally, the sharp increase in variance is attributed to the overlap reduction function oscillating around zero.

With the optimal estimator and its corresponding variance determined, we perform Bayesian inference using \Bilby~\cite{bilby_paper}.
For reasonable simplification, we do not consider the individual parameters of the BBH mass distribution.
We use $\langle \mathcal{M}_{c}^{5/3} \rangle$ as the representative mass parameter. 
It should be noticed that the simulation is based on GR, which is recovered in the limit $ \omega_{\rm BD}\rightarrow+\infty$.
To avoid explicitly dealing with $+\infty$, a practical approach is to rewrite $ \omega_{\rm BD} $ as $1/\omega_{\rm BD}$. 
Since $1/\omega_{\rm BD}$ spans multiple orders of magnitude, this parameter can be effectively transformed into logarithmic space. 
A uniform prior on $\mathrm{Log}_{10}(1/\omega_{\rm BD})$ assigns equal weight across different orders of magnitude, allowing us to better identify the scale at which BD gravity can be constrained.
Additionally, we impose a cutoff value of $\mathrm{Log}_{10}(1/\omega_{\rm BD}) = -10$ as a representation of GR.
Then the parameter space of $\boldsymbol{\Theta}$ becomes $\boldsymbol{\Theta}=\{\langle\mathcal{M}_{c}^{5/3}\rangle, \mathrm{Log}_{10}(1/\omega_{\rm BD}), \ \mathcal{R}^{\mathrm{BBH}}_{0}, \ \alpha_z, \ \beta_z, \ z_{p} \}$. 

\begin{figure}[htbp]
\centering
\includegraphics[width=0.8\textwidth]{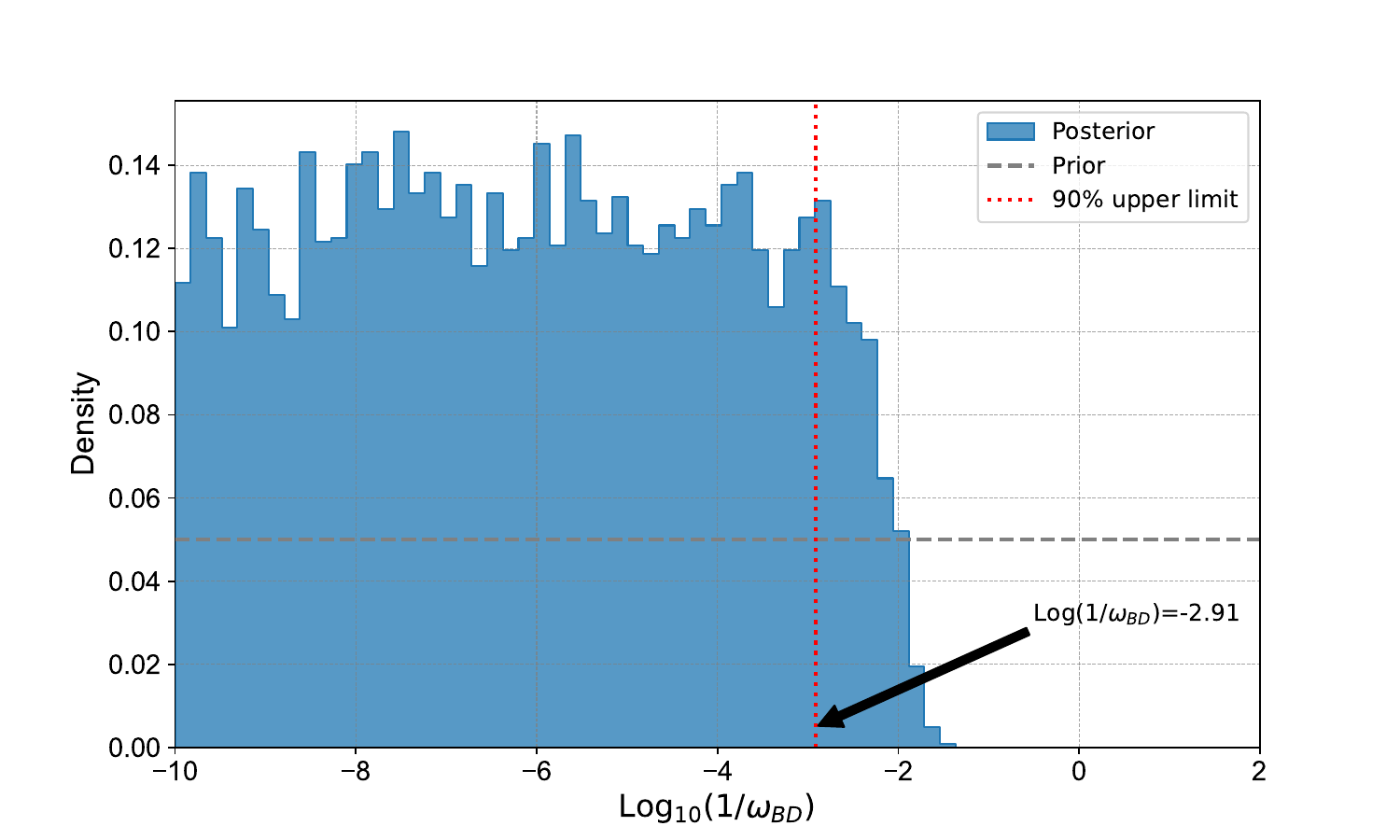}
\caption{
The posterior distribution of $\mathrm{Log}_{10}(1/\omega_{\rm BD})$, assuming that the population properties of BBH are well-constrained and treated as constants.
The prior distribution is denoted by the dashed line (grey).
The cutoff value of $\mathrm{Log}_{10}(1/\omega_{\rm BD})$ is chosen to be $-10$, corresponding to the recovery of GR.
The dotted line (red) indicate the upper limits $\mathrm{Log}_{10}(1/\omega_{\rm BD}) = -2.91$ at 90\% CL, which implies $\omega_{\rm BD} > 816.24$.
This constraint is considered highly idealized. 
More realistic constraints should account for the uncertainty in the population properties of BBH, as illustrated in Fig.~\ref{fig:all_parameters_pe.pdf}.
}
\label{fig:log_inv_w.png}
\end{figure}

For the result, we first consider an idealized scenario where all parameters related to the BBH population are fixed for the Bayesian analysis, leaving only the BD theory parameter $\omega_{\rm BD}$ as a free variable.
We choose a uniform prior on $\mathrm{Log}_{10}(1/\omega_{\rm BD})$ at the range of $[-10, 10]$.
Fig.~\ref{fig:log_inv_w.png} shows the posterior distribution of $\mathrm{Log}_{10}(1/\omega_{\rm BD})$.
The red dotted line denotes the upper limit of $\mathrm{Log}_{10}(1/\omega_{\rm BD})$ at 90\% credible level.
Converting it to $\omega_{\rm BD}$, we obtain the upper limit is $\omega_{\rm BD}>816.24$ at 90\% credible level.
It should be pointed that the upper limit demonstrates a certain dependence on the cutoff value of the prior for $\mathrm{Log}_{10}(1/\omega_{\rm BD})$.
However, a smaller cut-off value for the prior leads to a larger upper limit.

\begin{figure}[t]
\centering
\includegraphics[width=0.8\textwidth]{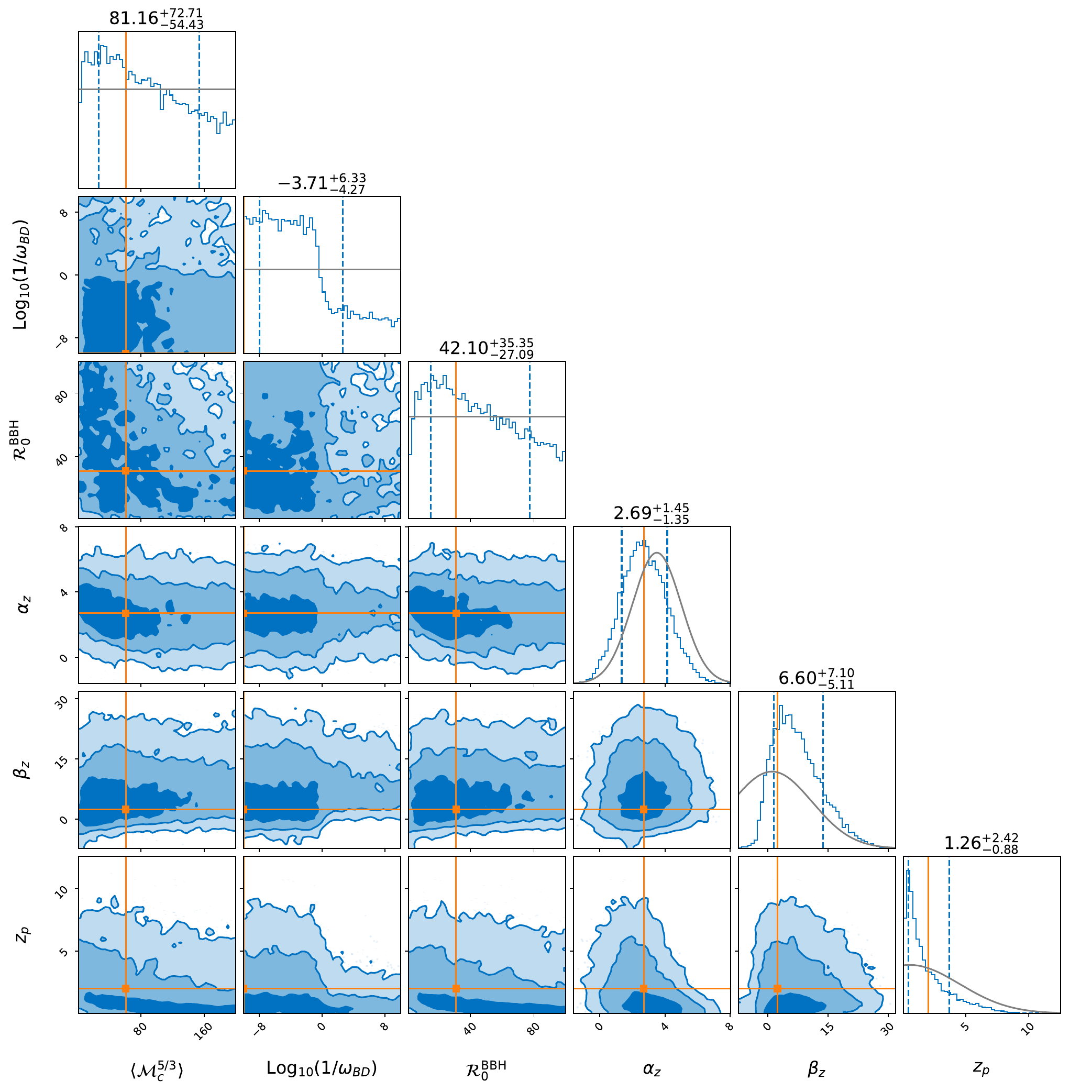}
\caption{
Full posterior distributions obtained by simultaneously inferring the coupling coefficient $\mathrm{Log}_{10}(1/\omega_{\rm BD})$ alongside the population parameters of BBH, i.e., the average mass and the merger rate of BBH.
Contour plots depict the $1\sigma$, $2\sigma$, and $3\sigma$ credible regions (blue, light blue, and very light blue, respectively).
The grey solid lines in each marginalized posterior illustrate the prior distribution for each parameter.
The vertical dashed blue lines represent the $1\sigma$ confidence interval. 
The injected values, denoted by the orange lines, are recovered within approximately $1\sigma$. Compared to the constraints on the BBH population parameters, the constraint on the BD parameter is relatively weak. 
}
\label{fig:all_parameters_pe.pdf}
\end{figure}

We then consider the full parameter space including the BBH population model. 
We adopt slightly narrower priors, aligning with measurements derived from direct detection of binary black hole mergers, compared to those specified in Ref.~\cite{Callister:2023tws}. 
These priors are listed in Table \ref{tab:priors} in detail.
Fig.~\ref{fig:all_parameters_pe.pdf} shows the posterior distributions of the parameters.
We effectively recover our prior on the parameters.
Compared with the last scenario that estimates $\mathrm{Log}_{10}(1/\omega_{\rm BD})$ solely, this case does not provide a more stringent result.
In this case, although $\omega_{\rm BD} >\mathcal{O}(1)$ holds a high probability ($>1 \sigma $), the possibility of $\omega_{\rm BD} < \mathcal{O}(1)$ cannot be excluded.
This outcome arises primarily from the degeneracy between the BD gravity parameter and the population parameters, with variations in the latter exerting a more significant influence on the background intensity. 
Consequently, this implies that the constraints on BD theory through SGWB observation strongly depend on the determination of the population model.
Comparing the posterior and prior distributions, we find that $\beta_{\rm z}$ and $z_{\rm p}$ are better constrained, whereas the other parameters for the BBH population are weakly constrained. 
Fortunately, the direct detection of binary black hole mergers provides constraints on the mass distribution (e.g., $\langle\mathcal{M}_{c}^{5/3}\rangle$), the local merger rate $\mathcal{R}^{\mathrm{BBH}}_{0}$, and the leading slope parameter $\alpha_z$~\cite{KAGRA:2021duu,KAGRA:2021kbb,Callister:2023tgi,Callister:2020arv,Li:2023yyt,Li:2024jzi,Rinaldi:2023bbd,Guo:2024wwv}.
Future improvements in the constraints on these population parameters enable us to constrain BD gravity more tightly.

\subsection{Seaching for scalar mode}
\label{sec:Seaching for scalar mode}

In this subsection, we focus on the search for a potential scalar gravitational-wave background arising from BNS and NSBH mergers in BD gravity.
The feasibility of simultaneously searching for scalar and tensor modes in the SGWB arises from the distinct differences in their frequency responses across different detector baselines, which are further reflected in their respective overlap reduction functions.

As discussed in Sec.~\ref{sec:SGWB spectrum in BD gravity}, the spectral index of the scalar mode background spectrum remains identical to that of the tensor mode in BNS systems. 
However, the NSBH systems introduce deviations to the spectral index in scalar mode, accompanied by a frequency-dependent oscillatory behavior.
While the oscillatory behavior may facilitate the search for scalar modes, the scalar background signal is expected to be substantially weaker than the tensor background signal under realistic conditions.
Furthermore, previous studies have utilized the mutually orthogonal approximation between different harmonics to disregard similar oscillatory effects caused by eccentric orbits~\cite{Moore:2018kvz,Hu:2023oiu}.

Based on the discussion above, the second one-year simulation dataset (denoted by ``Dataset \uppercase\expandafter{\romannumeral2}") is generated by a power-law without considering the population properties,
\begin{equation}\label{powerlaw_omega}
\Omega_{\mathrm{GW}}^{\rm T}(f) = \Omega_{\mathrm{ref}}^{\mathrm{T}}\left(\frac{f}{f_{\mathrm{ref}}}\right)^{\alpha_{\mathrm{T}}},\quad
\xi\cdot\Omega_{\mathrm{GW}}^\mathrm{S}=\Omega_{\mathrm{ref}}^{\mathrm{S}}\left(\frac{f}{f_{\mathrm{ref}}}\right)^{\alpha_{\mathrm{S}}},
\end{equation}
where $\Omega_{\mathrm{ref}}^{\mathrm{T}}$ is the reference amplitudes of the tensor mode at the reference frequency $f_{\mathrm{ref}} = 25$ Hz.
As decribed in Eqs.\,(\ref{widehat_Omega}) and (\ref{estimator_expectation_value}), the expectation value of the scalar component of the estimator from cross-correlation  statistic is given by $\xi\cdot\Omega_{\mathrm{GW}}^\mathrm{S}$, where $\xi$ is a constant determined by the specific value of $\omega_{\rm BD}$.
Thus $\Omega_{\mathrm{ref}}^{\mathrm{S}}$ here represents the effective reference amplitude of the scalar mode at $f_{\mathrm{ref}} = 25$ Hz, with $\xi$ absorbed into the amplitude of $\Omega_{\mathrm{GW}}^\mathrm{S}$.
The spectral indices for the tensor and scalar modes are denoted by $\alpha_{\mathrm{T}}$ and $\alpha_{\mathrm{S}}$, respectively.

\begin{table*}[t] 
    \centering
    \renewcommand{\arraystretch}{1.3} 
    \scriptsize
    \caption{The tensor and scalar background energy density $\Omega_{\mathrm{GW}}(f)$ at 25 Hz for BBH, BNS, and NSBH systems, as well as the total background contributions in BD theory, are calculated based on the population model described in Sec.~\ref{sec:Population}.}
    \label{tab:background energy density}
    \begin{tabular}{c|c c|c c|c c}
        \hline
        \multicolumn{1}{c|}{} & \multicolumn{2}{c|}{$\omega_{\rm BD}=1$} & \multicolumn{2}{c|}{$\omega_{\rm BD}=10$} & \multicolumn{2}{c}{$\omega_{\rm BD}=10000$} \\
        \cline{2-7}
        & $\Omega_{\mathrm{ref}}^{\mathrm{T}}$(25 Hz) & $\Omega_{\mathrm{ref}}^{\mathrm{S}}$(25 Hz) & $\Omega_{\mathrm{ref}}^{\mathrm{T}}$(25 Hz) & $\Omega_{\mathrm{ref}}^{\mathrm{S}}$(25 Hz) & $\Omega_{\mathrm{ref}}^{\mathrm{T}}$(25 Hz) & $\Omega_{\mathrm{ref}}^{\mathrm{S}}$(25 Hz) \\
        \hline
        BBH & $4.10 \times 10^{-9}$ & 0 & $1.44 \times 10^{-9}$ & 0 & $1.18 \times 10^{-9}$ & 0 \\
        BNS & $3.86 \times 10^{-9}$ & $8.93 \times 10^{-12}$ & $1.31 \times 10^{-9}$ & $1.48 \times 10^{-13}$ & $1.07 \times 10^{-9}$ & $1.60 \times 10^{-19}$ \\
        NSBH & $9.34 \times 10^{-10}$ & $1.25 \times 10^{-11}$ & $3.46 \times 10^{-10}$ & $3.26 \times 10^{-13}$ & $2.87 \times 10^{-10}$ & $3.35 \times 10^{-19}$ \\
        Total & $8.89 \times 10^{-9}$ & $2.14 \times 10^{-11}$ & $3.10 \times 10^{-9}$ & $4.74 \times 10^{-13}$ & $2.54 \times 10^{-9}$ & $4.95 \times 10^{-19}$ \\
        \hline
    \end{tabular}
\end{table*}

To illustrate the distinction between the contributions of the scalar and tensor polarization to the AGWB, we compute the corresponding amplitudes for the tensor component, $\Omega_{\mathrm{GW}}^{\rm T}$, and the scalar component, $\xi\cdot\Omega_{\mathrm{GW}}^\mathrm{S}$, at $f_{\mathrm{ref}} = 25$ Hz, using selected values of $\omega_{\rm BD}$.
These calculations utilize the population models for BBH, BNS and NSBH systems described in Sec.\ref{sec:Population}. The corresponding results are presented in Table \ref{tab:background energy density}, which demonstrate that the scalar background is significantly weaker than the tensor background, making the scalar signal likely to be overshadowed.
In this work, we set $\omega_{\rm BD} = 1$ as an appropriate deviation to enhance the contribution of the scalar background. 
We begin by fitting the computed tensor and scalar background spectra using the power-law model introduced in Eq.\,(\ref{powerlaw_omega}).
For the tensor component, contributions from BBH, BNS, and NSBH systems result in $\Omega_{\mathrm{ref}}^{\mathrm{T}}=8.90 \times 10^{-9}$ with a corresponding spectral index of 
$\alpha_{\mathrm{T}} = 0.66$.
The fitted spectral index of the tensor component aligns with the prediction of GR, as the tensor background is dominated by contributions from BBH and BNS systems, which exhibit equal sensitivities of compact binaries.
In comparison, the scalar component, driven by contributions from BNS and NSBH systems, results in $\Omega_{\mathrm{ref}}^{\mathrm{S}}=2.11 \times 10^{-11}$ with a corresponding spectral index of 
$\alpha_{\mathrm{S}} = 0.45$.

\begin{table*}[t]
    \centering
    \scriptsize
    \caption{Parameters and their prior distributions used in the analyses of Dataset \uppercase\expandafter{\romannumeral2}.}
    \label{tab:dataset2}
    \resizebox{\textwidth}{!}{ 
    \begin{tabular}{c p{4cm} c c c c}
        \hline\hline
        \textbf{Parameter} & \textbf{Description} & \multicolumn{3}{c}{\textbf{Fiducial Values}} & \textbf{Prior} \\[2pt]
        & & \textbf{(Primary)} & \textbf{(Ideal)} & \textbf{(Null)} & \\[2pt]
        \hline
        \rule{0pt}{12pt}
        $\Omega_{\mathrm{ref}}^{\mathrm{T}}$ 
        & The reference tensor background’s amplitude.
        & $8.90 \times 10^{-9}$
        & $8.90 \times 10^{-9}$
        & $8.90 \times 10^{-9}$ 
        & Log-Uniform $[10^{-16}, 10^{-5}]$ \\
        $\alpha_{\mathrm{T}}$ 
        & The spectral index of tensor background.
        & $0.66$ 
        & $0.66$
        & $0.66$
        & Uniform $[-3, 3]$ \\
        $\Omega_{\mathrm{ref}}^{\mathrm{S}}$ 
        & The effective reference scalar background’s amplitude.
        & $2.11 \times 10^{-11}$
        & $8.90 \times 10^{-9}$
        & $0$ 
        & Log-Uniform $[10^{-16}, 10^{-5}]$ \\
        $\alpha_{\mathrm{S}}$ 
        & The spectral index of scalar background.
        & $0.45$ 
        & $0.45$  
        & $0$
        & Uniform $[-3, 3]$ \\
        \hline\hline
    \end{tabular}
    } 
\end{table*}

These fitted values are used as the fiducial injection parameters, forming the primary simulation dataset for our analysis. 
Additionally, as an end-to-end test of the entire system, we constructed two additional control datasets: the ideal and null datasets. 
In the ideal dataset, the effective reference scalar background amplitude is amplified to match the reference tensor background amplitude. 
In contrast, in the null dataset, the scalar background is set to zero.
The parameters used for generating the simulations are summarized in Table \ref{tab:dataset2}.

\begin{figure}[t]
\centering
\includegraphics[width=0.8\textwidth]{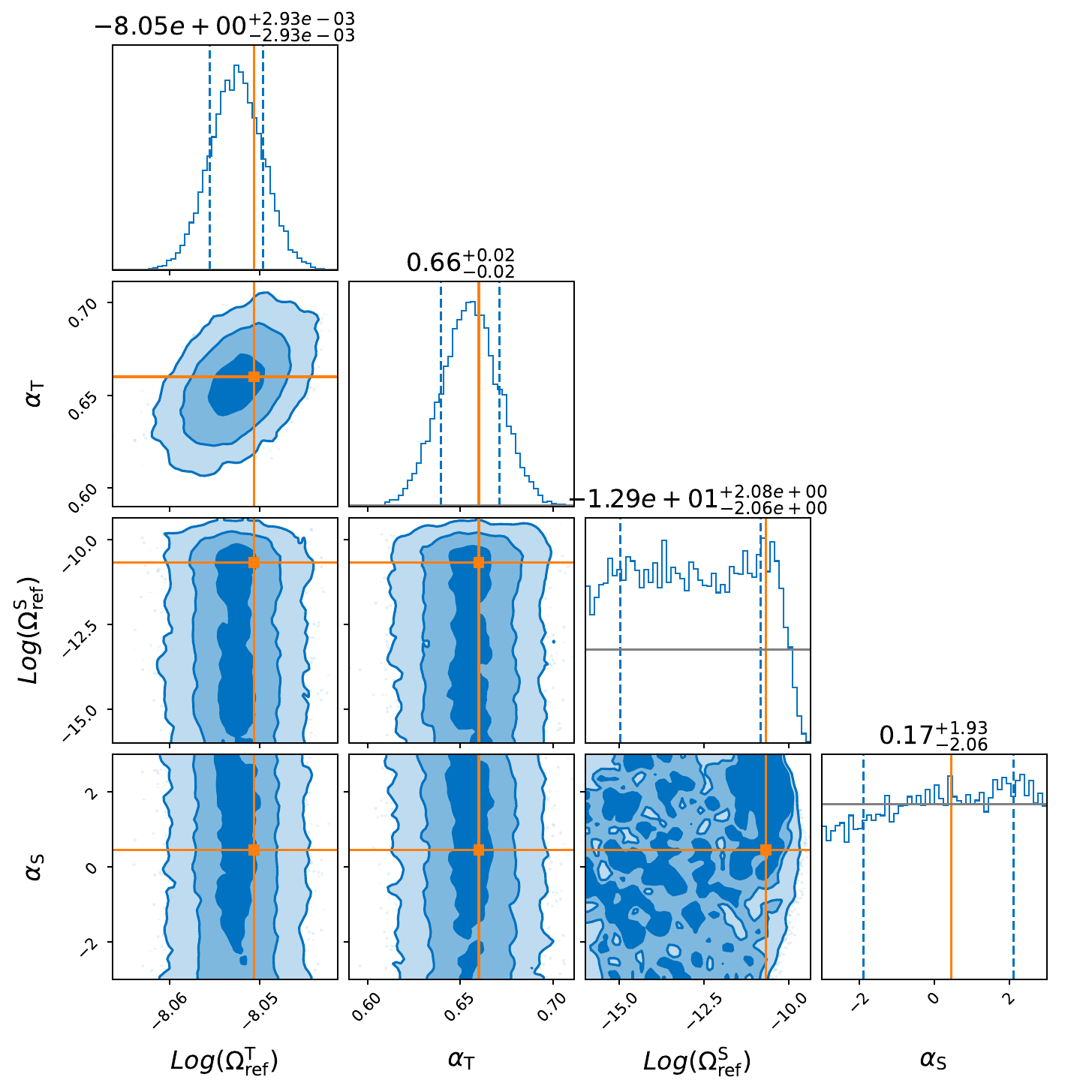}
\caption{
The results of parameter estimation for the primary simulation dataset in Dataset \uppercase\expandafter{\romannumeral2}.
Contour plots depict the $1\sigma$, $2\sigma$, and $3\sigma$ credible regions (blue, light blue, and very light blue, respectively).
The grey solid lines represent the prior distributions for each parameter.
The injected values, denoted by the orange lines.
The amplitude and seprctral index of tensor background are tightly constrained.
The upper limit on the scalar mode amplitudes of the background energy density is set at $\Omega_{\mathrm{ref}}^{\mathrm{S}} < 3.064 \times 10^{-11}$ (90\% CL).
The spectral index of the scalar mode of the background energy density is not significantly constrained.
}
\label{fig:omega_tensor_omega_scalar_primary.pdf}
\end{figure}

The corner plots for the primary simulation dataset are displayed in Fig.~\ref{fig:omega_tensor_omega_scalar_primary.pdf}, where we transform the posterior distributions of $\Omega_{\mathrm{ref}}^{\mathrm{T}}$ and $\Omega_{\mathrm{ref}}^{\mathrm{S}}$ into logarithmic space, denoted as $\mathrm{Log}_{10}\Omega_{\mathrm{ref}}^{\mathrm{T}}$ and $\mathrm{Log}_{10}\Omega_{\mathrm{ref}}^{\mathrm{S}}$, respectively. 
The amplitudes of the tensor mode $\Omega_{\mathrm{ref}}^{\mathrm{T}}$ and the corresponding spectral index $\alpha_{T}$ are recovered within 1$\sigma$ and are subject to stringent constraints. Specifically, the amplitude is determined to be $\Omega_{\mathrm{ref}}^{\mathrm{T}} = (8.91 \pm 0.06) \times 10^{-9}$, and the spectral index is $\alpha_{\mathrm{T}} = 0.66^{+0.02}_{-0.02}$ within the $1\sigma$ credible interval.
For the search of scalar gravitational waves, the upper limit on the amplitudes is determined to be $\Omega_{\mathrm{ref}}^{\mathrm{S}} < 3.12 \times 10^{-11}$ (90\% CL), corresponding to $\mathrm{Log}_{10}\Omega_{\mathrm{ref}}^{\mathrm{S}} < -10.51$ (90\% CL), while the spectral index $\alpha_{S}$ remains unconstrained.

\begin{figure}[htbp]
\centering
\includegraphics[width=.47\textwidth]{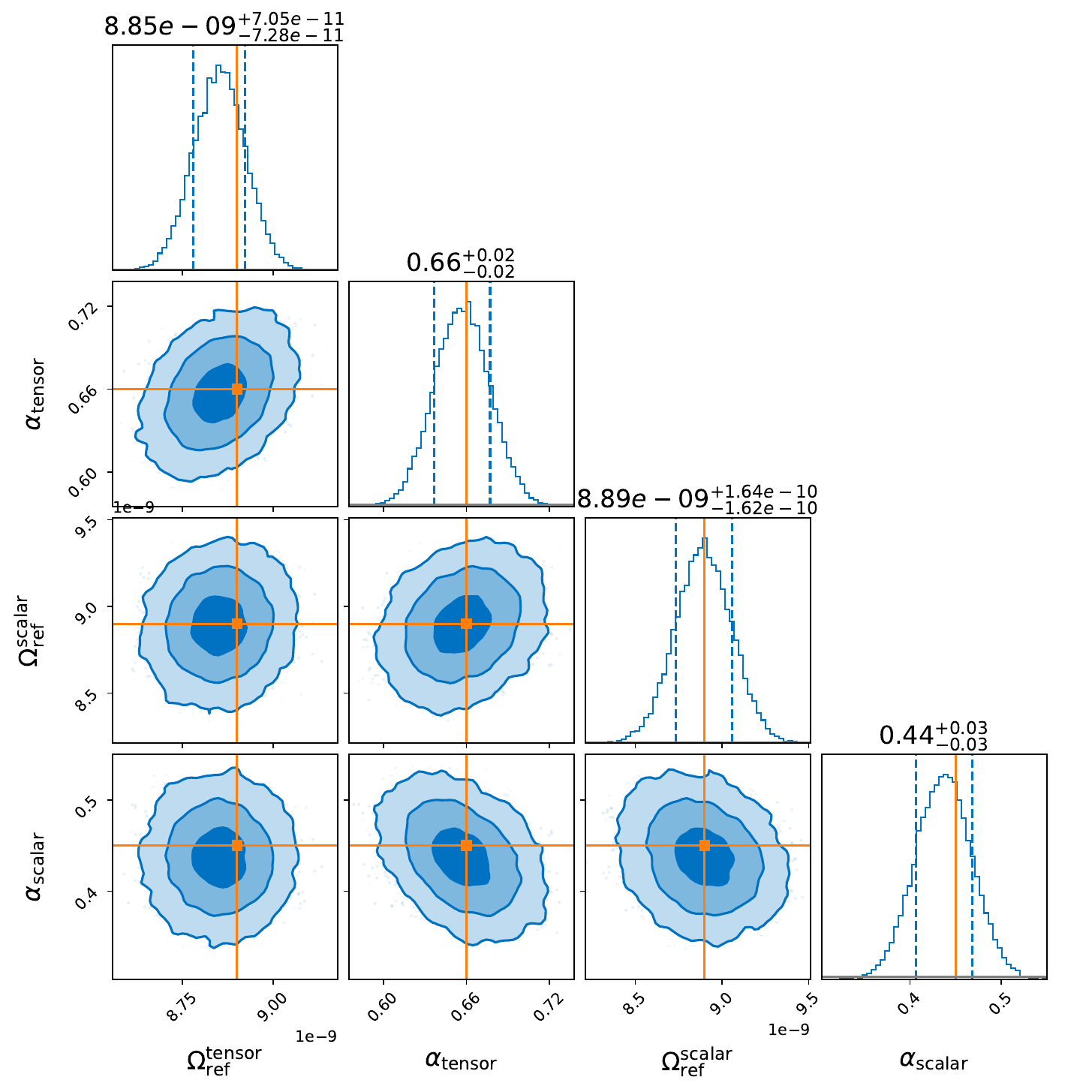}
\qquad
\includegraphics[width=.47\textwidth]{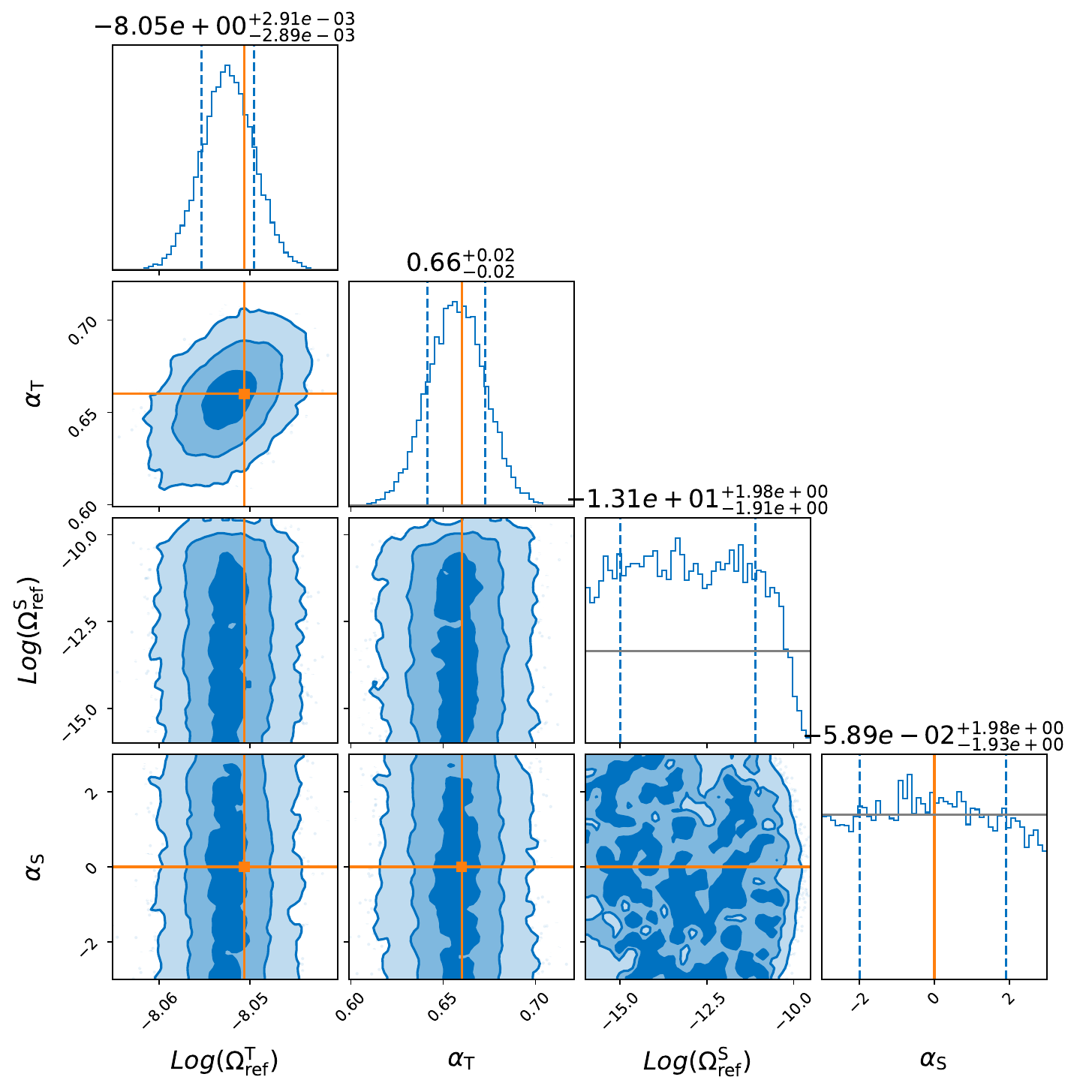}
\caption{
The results of parameter estimation for the ideal (left) and null (right) simulation dataset in Dataset \uppercase\expandafter{\romannumeral2}.
Contour plots depict the $1\sigma$, $2\sigma$, and $3\sigma$ credible regions (blue, light blue, and very light blue, respectively).
The grey solid lines represent the prior distributions for each parameter.
The injected values, denoted by the orange lines.
For the ideal dataset (left), strong constraints are placed on both tensor and scalar backgrounds. 
In contrast, a comparison of the results from null dataset (right) with those of the primary dataset indicates that the absence of a scalar background cannot be excluded.
}
\label{fig:ideal and null test}
\end{figure}

The weaker constraints on the scalar background, compared to the tensor background, arise from the significantly smaller contribution of the scalar mode to the overall background intensity, resulting in it being overshadowed by the tensor mode. 
This is further demonstrated by the ideal dataset, as shown on the left side of Fig.~\ref{fig:ideal and null test}, which serves as a control. 
When the scalar background intensity is amplified to match that of the tensor background, strong constraints can be imposed simultaneously on both scalar and tensor components. 
Specifically, the amplitudes are constrained to $\Omega_{\mathrm{ref}}^{\mathrm{T}} = (8.85 \pm 0.07) \times 10^{-9}$ for the tensor background and $\Omega_{\mathrm{ref}}^{\mathrm{S}} = (8.89 \pm 0.16) \times 10^{-9}$ for the scalar background. 
The corresponding spectral indices are $\alpha_{\mathrm{T}} = 0.66^{+0.02}_{-0.02}$ and $\alpha_{\mathrm{S}} = 0.44^{+0.03}_{-0.03}$. 
These results demonstrate that the search for tensor and scalar backgrounds enables the effective separation of mixed tensor and scalar polarization modes with comparable sensitivity to each mode.

Additionally, on the right side of Fig.~\ref{fig:ideal and null test}, we present the results for another control dataset, the null dataset. 
Comparing the posterior distribution of the null dataset to that in the primary dataset reveals that the tensor background is subject to equally stringent constraints in both cases.
A smaller upper limit is placed on the scalar background amplitude, $\Omega_{\mathrm{ref}}^{\mathrm{S}} < 1.85 \times 10^{-11}$ (90\% CL), while the spectral index of the scalar background still remains unconstrained.
The results above indicate that there is no evidence to exclude the absence of a scalar background, even in scenarios where BD gravity exhibits significant deviations from GR ($\omega_{\rm BD} = 1$), as the scalar background remains considerably weaker than the tensor background.

\section{Conclusions}\label{sec:Conclutions}

We take Brans-Dicke gravity as an example to comprehensively analyze how the generation and propagation effects of gravitational waveforms under scalar-tensor gravity are encoded into the stochastic gravitational wave background. 
Unlike previous works that directly utilized power-law models to search for the stochastic gravitational wave background, or employed Fisher information matrices for parameter estimation, we perform an end-to-end analysis of BD gravity with third-generation gravitational wave detectors. 
Utilizing realistic population properties of compact binary systems, we generate two sets of one-year astrophysical stochastic gravitational wave background simulation datasets and conducted a complete Bayesian analysis.

The first set of simulations demonstrates that, the uncertainty of the population properties has a more obvious impact on the stochastic gravitational wave background spectrum than the BD gravity, making it difficult to constrain the theory under such uncertainty. 
Even with perfect knowledge of the BBH population, the upper limit on the coupling constant $\omega_{\rm BD}$ only suggests $\omega_{\rm BD} > 816$.
The accurate measurements of the binary population from an increased number of observed systems in third-generation gravitational wave detectors~\cite{Singh:2021zah}, along with extended observation times of the AGWBs, can achieve more stringent constraints on BD gravity. 
Additionally, by matching waveforms and signals from a larger number of observed binary systems with high SNR,
these constraints will be further tightened~\cite{Punturo:2010zz,Zhang:2017sym}.

The second set of simulations indicates that one year of stochastic gravitational wave observations will yield strong constraints on the tensor mode background, while only upper limits can be placed on the scalar mode background, as even in scenarios with significant deviations from General Relativity, the scalar background is still expected to remain smaller than the tensor mode background.
Besides, the method shows great feasibility to search for scalar modes in the stochastic gravitational wave background. 
Longer observation times and larger detector networks with more baselines may offer opportunities to detect potential scalar modes.

In addition, although the tensor background contributions from NSBHs are relatively minor, the scalar background generated by NSBHs significantly modifies the spectral index of the background spectrum and introduces oscillatory features. These distinctive characteristics could aid in distinguishing this signal from GR predictions in the data, but detection remains challenging with ground-based gravitational wave detectors.
Future space-based gravitational wave detectors, such as the LISA-TianQin networks, are expected to detect alternative polarizations of stochastic gravitational wave backgrounds~\cite{Hu:2022byd} and may enhance detection prospects by capitalizing on the more rapidly oscillatory behavior of scalar background at low frequencies.
 
Moreover, utilizing noise-free correlation measurements from pulsar timing arrays will enhance our understanding of the nanohertz stochastic gravitational wave background, thereby providing deeper insights into gravity~\cite{Bernardo:2023pwt,Yunes:2024lzm}.
Observations of the stochastic gravitational wave background across a broader range of frequencies will undoubtedly yield more reliable information about gravity.


\acknowledgments
We thank all \Pygwb \ code authors for their patience in addressing our questions regarding the use of the code. 
We also thank Enrico Barausse, Shaopeng Tang, Yuanzhu Wang, Aoxiang Jiang, Bo Gao, Chi Zhang, Xingjiang Zhu, and Xiao Guo for their valuable discussions and comments.
This work is supported by the Natural Science Foundation of China (No. 12233011).
R.C is supported by China Scholarship Council, No.202406340069.
Z.L is supported by China Scholarship Council, No. 202306340128.
W.Z is supported by the National Key R\&D Program of China (Grant No. 2022YFC2204602 and 2021YFC2203102), Strategic Priority Research Program of the Chinese Academy of Science (Grant No. XDB0550300), the National Natural Science Foundation of China (Grant No. 12325301 and 12273035), the Fundamental Research Funds for the Central Universities (Grant No. WK2030000036 and WK3440000004), the Science Research Grants from the China Manned Space Project (Grant No.CMS-CSST-2021-B01), the 111 Project for "Observational and Theoretical Research on Dark Matter and Dark Energy" (Grant No. B23042).




\bibliographystyle{JHEP}
\bibliography{biblio.bib}

\end{document}